\documentclass[usenatbib]{mn2e}
\usepackage{times}
\usepackage{epsfig}

\newcommand{\be}{\begin{equation}}
\newcommand{\ee}{\end{equation}}
\newcommand{\etal}{et al.}

\newcommand{\mean}[1]{\left\langle#1\right\rangle}

\newcommand{\dm}{{\Delta}m_{12}}
\newcommand{\fsub}{f_{\rm sub}}
\newcommand{\kpc}{\rm kpc}
\newcommand{\Mpc}{\rm Mpc}
\newcommand{\Myr}{\rm Myr}
\newcommand{\hMpc}{h^{-1}{\rm Mpc}}
\newcommand{\keV}{\rm keV}
\newcommand{\pixel}{\rm pixel}
\newcommand{\Mvir}{M_{\rm virial}}

\newcommand{\mags}{{\rm mag}}
\newcommand{\ergs}{{\rm erg\,s^{-1}}}
\newcommand{\Msolpyr}{{\rm M_\odot\,yr^{-1}}}
\newcommand{\mJy}{{\rm mJy}}

\newcommand{\SFR}{{\rm SFR}}
\newcommand{\Gyr}{{\rm Gyr}}
\newcommand{\Om}{\Omega_M}
\newcommand{\Ol}{\Omega_\Lambda}

\newcommand{\Lxz}{L_{X,z}}
\def\Msol{\mathrel{\rm M_{\odot}}}
\def\ls{\mathrel{\hbox{\rlap{\hbox{\lower4pt\hbox{$\sim$}}}\hbox{$<$}}}}
\def\gs{\mathrel{\hbox{\rlap{\hbox{\lower4pt\hbox{$\sim$}}}\hbox{$>$}}}}
\title[LoCuSS: Brightest Cluster Galaxy Dominance at z=0.2] {LoCuSS:
  Connecting the Dominance and Shape of Brightest Cluster Galaxies
  with the Assembly History of Massive Clusters}

\author[Smith G.\ P., et al.]
       {Graham P.\ Smith,$\!^{1,2\star}$
        Habib G.\ Khosroshahi,$\!^{3,4\dagger}$
        A.\ Dariush,$\!^{1,5}$
        A.\ J.\ R.\ Sanderson,$\!^1$\newauthor
	T.\ J.\ Ponman,$\!^1$
	J.\ P.\ Stott,$\!^3$
        C.\ P.\ Haines,$\!^1$
        E.\ Egami,$\!^6$
	D.\ P.\ Stark$^7$
\vspace{1mm}\\ 
        $^1$ School of Physics and Astronomy, University of Birmingham, 
             Edgbaston, Birmingham, B15 2TT, England\\ 
        $^2$ California Institute of Technology, Mail Code 105--24, 
             Pasadena, CA~91125, USA\\
        $^3$ School of Astronomy, Institute for Studies in Theoretical 
             Physics and Mathematics (IPM), Tehran, Iran\\
        $^4$ Astrophysics Research Institute, Liverpool JMU, 
             Twelve Quays House, Egerton Wharf, Birkenhead, CH41 1LD, 
             England\\
	$^5$ Cardiff School of Physics and Astronomy, Cardiff University, 
             Queens Buildings, The Parade, Cardiff, CF24 3AA, Wales\\
        $^6$ Steward Observatory, University of Arizona, 933 North
             Cherry Avenue, Tucson, AZ 85721, USA \\
        $^7$ Institute of Astronomy, University of Cambridge,
             Madingley Road, Cambridge, CB3 0HA, England\\
        $^\star$Email: gps@star.sr.bham.ac.uk\\
        $^\dagger$Email: habib@ipm.ir}

\begin{document}

\date{Accepted, Received}

\pagerange{\pageref{firstpage}--\pageref{lastpage}} \pubyear{2009}

\maketitle

\label{firstpage}

\begin{abstract} 
  We study the luminosity gap, $\dm$, between the first and second
  ranked galaxies in a sample of 59 massive ($\sim10^{15}\Msol$)
  galaxy clusters, using data from the Hale Telescope, the
  \emph{Hubble Space Telescope (HST)}, \emph{Chandra}, and
  \emph{Spitzer}.  We find that $\dm$ distribution, $p(\dm)$, is a
  declining function of $\dm$, to which we fitted a straight line:
  $p(\dm)\propto-(0.13\pm0.02)\dm$.  The fraction of clusters with
  ``large'' luminosity gaps is $p(\dm\ge1)=0.37\pm0.08$, which
  represents a $3\sigma$ excess over that obtained from Monte Carlo
  simulations of a Schechter function that matches the mean cluster
  galaxy luminosity function.  We also identify four clusters with
  ``extreme'' luminosity gaps, $\dm\ge2$, giving a fraction of
  $p(\dm\ge2)=0.07^{+0.05}_{-0.03}$.  More generally, large luminosity
  gap clusters are relatively homogeneous, with elliptical/disky
  brightest cluster galaxies (BCGs), cuspy gas density profiles (i.e.\
  strong cool cores), high concentrations, and low substructure
  fractions.  In contrast, small luminosity gap clusters are
  heterogeneous, spanning the full range of boxy/elliptical/disky BCG
  morphologies, the full range of cool core strengths and dark matter
  concentrations, and have large substructure fractions.  Taken
  together, these results imply that the amplitude of the luminosity
  gap is a function of both the formation epoch, and the recent infall
  history of the cluster.  ``BCG dominance'' is therefore a phase that
  a cluster may evolve through, and is not an evolutionary
  ``cul-de-sac''.  We also compare our results with semi-analytic
  model predictions based on the Millennium Simulation.  None of the
  models are able to reproduce all of the observational results on
  $\dm$, underlining the inability of the current generation of models
  to match the empirical properties of BCGs.  We identify the strength
  of AGN feedback and the efficiency with which cluster galaxies are
  replenished after they merge with the BCG in each model as possible
  causes of these discrepancies.
\end{abstract}


\section{Introduction}\label{sec:intro}

Numerical simulations and large-scale redshift surveys both indicate
that we live in a hierarchical universe, i.e.\ one in which the
large-scale structure of the universe grows from the bottom up with
smaller objects forming earlier than larger objects.  This picture
rests on the matter content of the universe being dominated by
collisionless dark matter particles, smoothly distributed at early
times, and seeded with small density perturbations.  Exploring this
picture observationally in the non-linear regime of gravitational
collapse, i.e.\ within collapsed dark matter halos that host
individual galaxies through to massive clusters of galaxies,
complements the statistical analysis of the linear regime probed by
galaxy redshift surveys.  Indeed, a promising route to fleshing out
our understanding of hierarchical structure formation is to measure
observable quantities that are sensitive to the age and/or assembly
history of dark matter halos, and thus in principle to test the
hierarchical paradigm by comparing the observed and predicted
distributions.  Any discrepancies found between observation and theory
may ultimately point to modifications to the theoretical model
including, for example, the properties of the dark matter particle and
the distribution of initial density fluctuations
\citep[e.g.][]{komatsu09}.  Quantities discussed in the literature
that may be useful probes of the age and assembly history of dark
matter halos include the luminosity gap between the first and second
ranked galaxies in a group or cluster (often expressed as the
difference between their magnitudes, $\dm=m_1-m_2$; e.g.\
\citealt{dariush07}), the concentration of dark matter halos
\citep[e.g.][]{neto07,okabe09}, and the sub-halo population of dark
matter halos \citep[e.g.][]{taylor04,zentner05}.

The luminosities of the first and second ranked galaxies in clusters
was first studied, as far as we are aware, by \citet[][see also
\citealt{Geller76,Tremaine77,Ostriker77,Oergerle89}]{sandage73}.  More
recently the luminosity gap, $\dm$, was studied in the context of
galaxy groups, the term ``fossil'' being coined to describe virialized
systems with $\dm\ge2$ \citep{ponman94,jones00,jones03}.  $L^\star$
galaxies are absent from fossil groups, which were thus interpreted as
having formed at early times, with dynamical friction then having
sufficient time to cause the $L^\star$ galaxy population to merge and
form the brightest group galaxy (BGG).  Fossil groups are expected to
be more common than fossil clusters, at least in part because the
probability of galaxy-galaxy merging is anti-correlated with galaxy
velocity, and thus with cluster mass.  Nevertheless, two clusters with
masses of $\sim10^{14}\Msol$ have been found with $\dm>2$
\citep{kmpj06,mendes06,cyp06}.  These two objects (RX\,J1416.4$+$2315,
and RX\,J15552.2$+$2013) raise the interesting question of whether the
most massive (${\gs}10^{15}\Msol$) clusters might host similarly
dominant BCGs.  Theoretical studies suggest that this is the case, for
example, \cite{milos06} and \cite{dariush07} predict that $\sim1-3\%$
and $\sim5\%$ of $10^{15}\Msol$ clusters have $\dm\ge2$ respectively.

The concentration parameter of a dark matter halo describes the shape
of its density profile following the so-called universal profile
proposed by \citet{nfw97} and variants thereon.  Halos with smaller
concentrations have a flatter density profile, while larger
concentrations imply a steeper density profile.  \cite{bullock01}
analyzed numerical simulations of CDM universes finding a weak
dependence of concentration on halo mass: $c\propto M^\alpha$ with
$\alpha\simeq-0.1$ \citep[see also][]{Dolag04,neto07,duffy08}.  This
relationship arises from the relative timing of the formation of dark
matter halos as a function of mass.  On average less massive halos
form at earlier times than more massive halos in a hierarchical
universe.  At earlier times the universe was denser than at later
times, and thus the central regions of less massive halos are
relatively dense, leading to higher concentration parameters than for
more massive halos.  Some observational studies have reported very
high concentration parameters in individual systems, for example, one
of the fossil groups studied by \cite{kjp04,kmpj06} was found to have
$c>50$, based on modeling of X-ray observations.  Lensing studies of
several individual strong-lensing clusters have also obtained very
high concentrations of $c\sim10$ in contrast to the theoretical
prediction of $c\sim5$
\citep{kneib03,gavazzi03,broadhurst05,limousin07}.  More recently,
observational studies have begun to study larger samples and thus to
constrain the concentration-mass relation itself, and to probe the
general population rather than a small number of potentially extreme
objects \citep[e.g.][]{buote07,okabe09}.

Theoretically, substructures within dark matter halos, i.e.\ the
sub-halo population, are also sensitive to the assembly history of the
host dark matter halo \citep[e.g.][]{taylor04,zentner05}.
Observationally substructures in galaxy clusters can be identified via
detailed modeling of the observed gravitational lensing signal
\citep{smith05,smith09,richard09,richard10}.  Specifically, group- and
galaxy-scale perturbers are required to achieve statistically
acceptable fits to the strong-lensing data.  The contribution of these
structures to the total cluster mass is quantified via the
``substructure fraction'', $\fsub$, defined as the amount of mass
within the adopted cluster-centric radius that is assigned to
substructures divided by the total cluster mass within the same
aperture.  \cite{smith08} combined \citeauthor{smith05}'s
(\citeyear{smith05}) observational measurements of $\fsub$ for 10
X-ray luminous galaxy clusters with \citeauthor{taylor04}'s
(\citeyear{taylor04}) semi-analytic model of structure formation to
explore the interpretation of lensing-based measurements of $\fsub$.
The main conclusion was that $\fsub$ depends on both when the cluster
formed, and on the level of recent mass assembly, each defined as the
lookback time to when each cluster had acquired 50\% and 90\% of their
observed mass respectively.  For example: clusters at $z=0.2$ with
$\fsub<0.1$ formed on average at $z\gs0.8$, and suffered $\le10\%$
mass growth since $z=0.4$; in contrast, clusters at $z=0.2$ with the
highest substructure fractions ($\fsub\gs0.4$) formed on average since
$z\simeq0.4$ and acquired $\gs10\%$ of their mass between $z=0.25$ and
$z=0.2$, i.e.\ a time interval of just $500\Myr$.  

A complementary view of hierarchical merging within galaxy clusters is
available from BCG morphology.  Based on their isophotal shapes,
elliptical galaxies have been classified as disky or boxy
\citep{bender89} -- with positive and negative fourth-order Fourier
coefficients respectively.  The interpretation of boxy and disky
isophotes in terms of the details of galaxy merger histories is a
controversial subject \citep{faber97,naab03,khochfar05}.  In this
study we will side-step these difficulties, and concentrate simply on
disky/boxy isophotes as an indicator of the presence of gas that has
dissipated, and settled into a disk-like structure, either because the
last massive galaxy to merge with the BCG was gas rich, or because gas
has been accreted by the merger product from its environment, e.g.\ by
a BCG in a cool core cluster.  BCGs are the most massive early-type
galaxies and are generally expected to have boxy isophotes consistent
with formation via mergers of early-type (gas-poor) galaxies
\citep{lin04b,kpj06}.  However, BGGs in some fossil groups are as
bright as BCGs and do \emph{not} have boxy isophotes \citep{kpj06},
suggesting that (i) fossil BGGs may form early from the mergers
between gas-rich spiral galaxies, and (ii) some fossil BGGs may have
subsequently evolved into BCGs.

The main aim of this article is to combine measurements of the
luminosity gap, cool core strength, concentration, substructure
fraction, and BCG isophotal shape for a large sample of clusters to
assemble an empirical picture of the hierarchical assembly of clusters
and their BCGs.  We present the first observational measurement of the
distribution of the luminosity gap statistic of $10^{15}\Msol$
clusters, and compare this distribution with the other probes of
hierarchical assembly discussed above.  This allows us to build an
empirical picture with which to assess the usefulness of the
respective measurements for assessing the age of clusters.  We also
investigate how well the current generations of galaxy formation
models can reproduce the observed luminosity gap distribution.  The
data used for this study are drawn from the Local Cluster Substructure
Survey (LoCuSS; PI: Smith; http://www.sr.bham.ac.uk/locuss).  A
summary of LoCuSS is provided in \S\ref{sec:data}, together with a
description of the data used in this paper.  The analysis and results
are then presented in \S\ref{sec:analysis}, and compared with
theoretical predictions in \S\ref{sec:theory}.  The main conclusions
are summarized and discussed in \S\ref{sec:conclusions}.  We assume
$H_0{=}70\,{\rm km\,s^{-1}\,Mpc^{-1}}$, $\Omega_M{=}0.3$ and
$\Omega_\Lambda{=}0.7$ throughout.  In this cosmology $1\arcsec$
corresponds to a physical scale of $3.3\kpc$, at $z{=}0.2$.  All
photometric measurements are relative to Vega.

\section{Data}\label{sec:data}

\subsection{Sample Selection and Observing Strategy}\label{sec:sample}

LoCuSS is a morphologically-unbiased multi-wavelength survey of X-ray
luminous galaxy clusters at $0.15\le z\le0.3$.  The overall aim of the
survey is to measure the cluster-cluster scatter in key observables
such as the X-ray temperature, and $Y_X$ parameter
\citep{smith05,zhang08,okabe10}, the Sunyaev-Zeldovich Effect $Y$
parameter \citep{marrone09}, and the obscured and unobscured star
formation activity
\citep{haines09a,haines09b,haines10,smith10,pereira10}, and to
correlate this scatter with the structure and thus hierarchical
assembly history of the clusters.  The backbone of the survey is the
gravitational lensing analysis of \emph{HST} \cite[][Hamilton-Morris
et al.\ in prep.; May et al.\ in prep.]{smith05,richard10} and Subaru
\citep{okabe09,Oguri10} imaging data, because the lensing-based mass
maps can be used to infer the likely assembly history of the clusters
\citep{smith08}.

\begin{figure}
\center
\epsfig{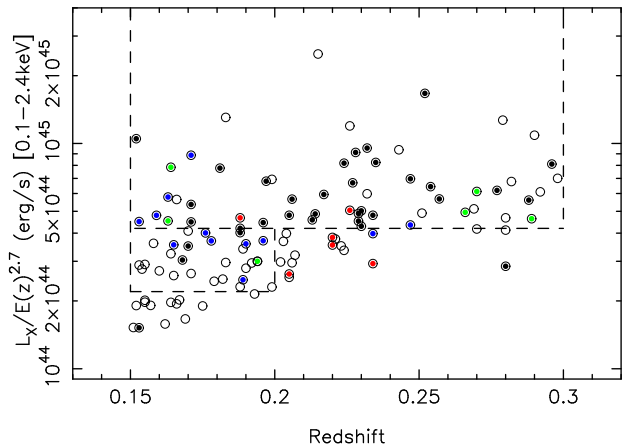}
\caption{ The distribution of the parent sample of 115 clusters in the
  $L_X-$redshift plane.  Filled data points indicate clusters that we
  observed in acceptable conditions (see \S\ref{sec:nirdata}) with
  WIRC on the Hale 200-in telescope, and are colour-coded as follows:
  black -- also observed with both \emph{HST} and \emph{Chandra}; blue
  -- also observed with \emph{HST}, but not with \emph{Chandra}; green
  -- also observed with \emph{Chandra}, but not with \emph{HST}; red
  -- observed with neither \emph{HST} nor \emph{Chandra}.  Open data
  points were not observed with WIRC at Palomar, and therefore do not
  form part of the sample studied in this article.  The dashed lines
  delineate the volume limited samples against which the observed
  sample is compared statistically in \S\ref{sec:tests}.  The absence
  of clusters from the parent sample to the lower right is caused by
  the flux-limit of the \emph{ROSAT} All-sky Survey. }
\label{fig:sample}
\end{figure}

The parent sample for this study comprises 115 clusters satisfying
$-27^\circ{\le}{\delta}{\le}70^\circ$, $0.15{\le}z{\le}0.3$,
$n_H{\le}7{\times}10^{20}{\rm cm}^{-2}$ drawn from the ROSAT All-sky
Survey catalogs \citep{ebeling98,ebeling00,bohringer04}.  The cut at
$\delta=-27^\circ$ ensures that the clusters are observable from
Palomar Observatory at elevations above $30^\circ$ (an airmass of
$\sec z\le2$).  The clusters span a decade in X-ray luminosity in the
$0.1-2.4\keV$ band: $2\times10^{44}\ls L_X\ls20\times10^{44}\ergs$
(Fig.~\ref{fig:sample}), which corresponds to a mass range of
$5\times10^{14}\ls\Mvir\ls3\times10^{15}\Msol$ \citep{reiprich02} --
i.e.\ well-matched to $10^{15}\Msol$.

The radius at which the mean enclosed density of a $10^{15}\Msol$ dark
matter halo at the median redshift of the cluster sample ($z=0.22$) is
$\mean{\rho(<r)}=200\rho_{\rm crit}$ is $r_{200}=1.9\Mpc$.  To ensure
that our results are comparable with previous studies of the
luminosity gap statistic, data that probe out to $\sim0.5r_{200}$
($\sim4.5\,{\rm arcmin}$) are required.  This requirement is met by
the Wide-field Infrared Camera (WIRC) on the Hale 200in Telescope at
Palomar Observatory (\S\ref{sec:nirdata}).  Traditionally, the
luminosity gap statistic has been studied at optical wavelengths.  In
contrast, working in the near-infrared permits the use of $(J-K)$
colours as a surrogate for a photometric redshift estimate of cluster
galaxies (\S\ref{sec:phot}), taking advantage of the relative
insensitivity of near-infrared colours to spectral type
\citep[e.g.][]{mannucci01}.  This is vital, in the absence of
exhaustive spectroscopic catalogs, to weed out non-cluster members
when calculating the luminosity gap statistic.

\subsection{Ground-based Near-infrared Data}\label{sec:nirdata}

\setcounter{footnote}{9}

The parent sample of 115 clusters were used as a back-up observing
program during observing runs with WIRC \citep{wilson03} on the Hale
200in Telescope\footnote{The Hale Telescope at Palomar Observatory is
  owned and operated by the California Institute of Technology.}
during observing runs spanning April 2004 to July 2005.  Data were
acquired when the full width half maximum (FWHM) of point sources
exceeded 1\,arcsec.  In total, data were obtained on 78 clusters, with
no pre-selection on cluster properties other than the X-ray selection
described above (\S\ref{sec:sample}).  Each cluster was observed with
a single $8.7'{\times}8.7'$ WIRC pointing.  BCGs at $z{\simeq}0.2$
have a typical angular extent of ${\sim}1{\rm arcmin}$; the individual
exposures were therefore dithered within a box of full-width $80''$ to
minimise inclusion of BCG flux in sky-flats constructed from the
science data.  Each cluster was observed for a total of 600\,sec per
filter, split into 5 dither positions.

\begin{figure}
\centerline{
  \epsfig{file=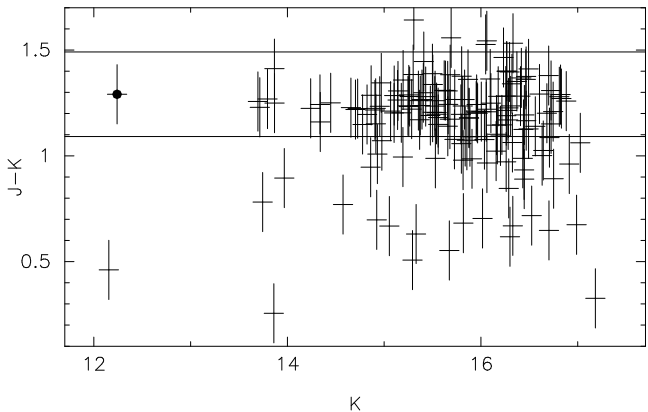,width=85mm}
}
\caption{The $(J-K)/K$ colour-magnitude relation for A\,1763. The
horizontal lines show the region within which likely cluster galaxies
were selected.  The filled circle denotes the BCG.}
\label{fig:cm}
\end{figure}

\begin{table*}
\begin{center}
\caption{The observed sample of clusters.\label{tab:sample}}
\begin{tabular}{lcccccl}
\hline
Cluster& ${\rm \alpha,\delta}$ [J2000] & Redshift  & $M_{K,BCG}$$^a$  &
${\dm}^a$ & HST PID & Also known as\\
\hline
A\,68  & 00 37 05.28 $+$09 09 10.8 & 0.255 & $-$26.65 & 0.25 & 8249 & \\ 
A\,115 & 00 55 50.65 $+$26 24 38.7 & 0.197 & $-$26.29 & 0.39 & 11312 & \\ 
A\,141 & 01 05 37.17 $-$24 40 49.7 & 0.230 & $-$26.54 & 0.40 & 10881 & RXC\,J0105.5$-$2439\\
ZwCl\,0104.4$+$0048 & 01 06 49.50 $+$01 03 22.1 & 0.255 & $-$26.37 & 0.52 & 11312 & Z\,348 \\
A\,209 & 01 31 53.00 $-$13 36 34.0 & 0.206 & $-$26.76 & 0.87 & 8249 & RXC\,J0131.8$-$1336 \\
A\,267 & 01 52 48.72 $+$01 01 08.4 & 0.230 & $-$26.56 & 1.43 & 8249 & RXCJ0152.7$+$0100\\
A\,291 & 02 01 43.11 $-$02 11 48.1 & 0.196 & $-$25.49 & 1.79 & 8301 & RXC\,J0201.7$-$0212\\
A\,383 & 02 48 02.00 $-$03 32 15.0 & 0.188 & $-$26.24 & 1.76 & 8249 & RXC\,J0248.0$-$0332\\
RXC\,J0331.1$-$2100 & 03 31 05.87 $-$21 00 32.7 & 0.188 & $-$27.07 & 0.96 & 10881 & \\
A\,521 & 04 54 06.88 $-$10 13 24.6 & 0.247 & $-$26.47 & 0.00 & 11312 & RXC\,J0454.1$-$1014 \\
A\,586 & 07 32 20.42 $+$31 37 58.8 & 0.171 & $-$26.31 & 0.51 & 8301 & \\
ZwCl\,0740.4$+$1740 & 07 43 23.16 $+$17 33 40.0 & 0.189 & $-$26.26 & 1.62 & 11312 & Z\,1432\\
A\,611 & 08 00 55.92 $+$36 03 39.6 & 0.288 & $-$26.90 & 1.28 & 9270 & \\
A\,665 & 08 30 57.36 $+$65 51 14.4 & 0.182 & $-$25.81 & 0.66 &  & \\
ZwCl\,0839.9$+$2937 & 08 42 56.06 $+$29 27 25.7 & 0.194 & $-$26.45 & 0.67 & 11312 & Z\,1883\\
ZwCl\,0857.9$+$2107 & 09 00 36.86 $+$20 53 40.0 & 0.235 & $-$25.82 & 0.34 & 8301 & Z\,2089\\
A\,750 & 09 09 12.74 $+$10 58 29.1 & 0.163 & $-$26.52 & 0.95 & 11312 & \\
A\,773 & 09 17 54.00 $+$51 42 57.6 & 0.217 & $-$26.68 & 0.47 & 8249 & \\
ZwCl\,0923.6$+$5340 & 09 27 10.69 $+$53 27 30.9 & 0.205 & $-$25.86 & 0.55 & & Z\,2379\\ 
ZwCl\,0949.6$+$5207 & 09 52 47.52 $+$51 53 27.6 & 0.214 & $-$26.45 & 1.24 & 8301 & Z\,2701\\
A\,901 & 09 56 26.40 $-$10 04 12.0 & 0.163 & $-$25.90 & 0.08 & 10395 & RXC\,J0956.4$-$1004\\
RX\,J1000.5$+$4409 & 10 00 31.16 $+$44 08 42.5 & 0.153 & $-$25.72 & 0.35 & 10881 & \\
A\,963 & 10 17 01.20 $+$39 01 44.4 & 0.205 & $-$26.42 & 1.73 & 8249 & \\
A\,1201 & 11 12 54.61 $+$13 26 08.2 & 0.169 & $-$26.25 & 1.37 & 8719 & \\
A\,1204 & 11 13 20.55 $+$17 35 39.1 & 0.171 & $-$25.80 & 0.95 & 8301 & \\
A\,1246 & 11 23 58.83 $+$21 28 45.4 & 0.190 & $-$25.92 & 0.42 & 8301 & RXCJ1123.9$+$2129\\
A\,1423 & 11 57 17.43 $+$33 36 38.6 & 0.213 & $-$26.16 & 1.77 & 8719 & \\
A\,1553 & 12 30 48.95 $+$10 32 45.6 & 0.165 & $-$26.82 & 0.94 & & \\ 
ZwCl\,1231.4$+$1007 & 12 34 17.45 $+$09 45 58.1 & 0.229 & $-$26.21 & 0.71 & 8719 & Z\,5247\\
A\,1634 & 12 54 01.84 $-$06 42 14.4 & 0.196 & $-$25.82 & 0.61 & & RXC\,J1254.0$-$0642\\ 
A\,1682 & 13 06 47.89 $+$46 33 32.5 & 0.226 & $-$26.96 & 0.09 & 8719 & \\
ZwCl\,1309.1$+$2216 & 13 11 46.15 $+$22 01 36.8 & 0.266 & $-$26.18 & 2.17 & & Z\,5768\\ 
A\,1704 & 13 14 24.38 $+$64 34 31.0 & 0.220 & $-$26.34 & 1.41 & & \\ 
A\,1758 & 13 32 44.47 $+$50 32 30.5 & 0.280 & $-$26.26 & 0.22 & & \\ 
A\,1763 & 13 35 16.32 $+$40 59 45.6 & 0.228 & $-$26.84 & 1.59 & 8249 & \\
A\,1835 & 14 01 02.40 $+$02 52 55.2 & 0.253 & $-$27.32 & 2.44 & 8249 & \\
A\,1914 & 14 25 59.78 $+$37 49 29.1 & 0.171 & $-$26.62 & 1.33 & 8301 & \\
A\,1961 & 14 44 31.85 $+$31 13 34.3 & 0.234 & $-$26.47 & 0.31 & & \\ 
A\,1994 & 14 56 13.48 $-$05 48 56.6 & 0.220 & $-$26.45 & 0.55 & & RXC\,J1456.3$-$0549\\ 
MS\,1455.0$+$2232 & 14 57 15.23 $+$22 20 34.0 & 0.258 & $-$26.06 & 0.11 & 8301 & ZwCl\,1454.8$+$2233, Z7160\\
A\,2009 & 15 00 19.63 $+$21 22 08.9 & 0.153 & $-$25.99 & 0.17 & 8301 & \\
ZwCl\,1459.4$+$4240 & 15 01 23.13 $+$42 20 39.6 & 0.290 & $-$26.28 & 0.07 & & Z\,7215\\ 
A\,2111 & 15 39 40.51 $+$34 25 27.0 & 0.229 & $-$25.58 & 0.37 & & \\ 
A\,2146 & 15 56 09.05 $+$66 21 33.1 & 0.234 & $-$26.07 & 0.41 & 8301 & \\
A\,2163 & 16 15 34.10 $-$06 07 26.0 & 0.169 & $-$25.46 & 0.49 & & \\ 
A\,2204 & 16 32 46.94 $+$05 34 31.3 & 0.152 & $-$25.82 & 0.11 & 8301 & \\
A\,2218 & 16 35 52.80 $+$66 12 50.4 & 0.171 & $-$26.12 & 0.32 & 5701 & \\
A\,2219 & 16 40 22.56 $+$46 42 21.6 & 0.228 & $-$26.62 & 1.18 & 6488 & \\
A\,2254 & 17 17 45.96 $+$19 40 48.0 & 0.178 & $-$26.05 & 0.88 & 8301 & \\
RX\,J1720.1$+$2638 & 17 20 10.14 $+$26 37 30.9 & 0.164 & $-$26.37 & 1.76 & 11312 & \\
A\,2261 & 17 22 27.24 $+$32 07 56.7 & 0.224 & $-$26.34 & 2.32 & 8301 & \\
RXC\,J2102.1$-$2431 & 21 02 09.98 $-$24 32 01.8 & 0.188 & $-$26.82 & 2.04 & & \\ 
A\,2345 & 21 27 13.73 $-$12 09 46.1 & 0.176 & $-$26.67 & 1.09 & 11312 & RXC\,J2127.1$-$1209\\
RX\,J2129.6$+$0005 & 21 29 40.02 $+$00 05 20.9 & 0.235 & $-$26.78 & 1.93 & 8301 & \\
A\,2390 & 21 53 36.72 $+$17 41 31.2 & 0.233 & $-$26.21 & 1.50 & 5352 & \\
RXC\,J2211.7$-$0350 & 22 11 45.95 $-$03 49 45.3 & 0.270 & $-$26.35 & 1.71 & & \\ 
A\,2485 & 22 48 31.13 $-$16 06 25.6 & 0.247 & $-$26.59 & 0.00 & 11312 & RXC\,J2248.5$-$1606\\
A\,2537 & 23 08 23.20 $-$02 11 31.0 & 0.297 & $-$26.19 & 0.63 & 9270 & RXC\,J2308.3$-$2011\\
A\,2631 & 23 37 39.82 $+$00 16 16.9 & 0.278 & $-$26.35 & 0.64 & 11312 & RXC\,J2337.6$+$0016\\
\hline
\end{tabular}
\end{center}
$^a$~Uncertainties on $M_{K,BCG}$ and $\dm$ are dominated by the
uncertainties on the photometric calibration, which is $\sim0.1$\,mag
in J- and K-bands. 
\end{table*}

The data were reduced in a uniform and standard manner using an
automated pipeline of {\sc iraf} tasks to dark subtract, flat-field,
align, and co-add the individual frames at the telescope.  Data
acquired in conditions worse than ${\rm FWHM}=1.5''$ suffered strongly
variable transparency and/or non-uniform background, and were
therefore excluded from the analysis, leaving a total of 59 clusters
with good quality data (Table~\ref{tab:sample}).  Astrometric and
photometric calibration were achieved by reference to the 2MASS
catalogs, to root mean square (rms) precisions of 1\,arcsec and
$0.1$\,magnitudes respectively \citep{stott08}.  The results described
in this article are insensitive to the uncertainty on the photometric
calibration.  An example $(J{-}K)/K$ colour magnitude diagram is shown
in Fig.~\ref{fig:cm}.  The typical depth reached by the data is
$K{\simeq}17$; an $L^\star$ galaxy has $K{\simeq}15$ and a typical BCG
has $K{\sim}12{-}13$ at $z{\simeq}0.2$.

\subsection{\emph{Hubble Space Telescope} Observations}\label{sec:hst}

\emph{Hubble Space Telescope (HST)}\footnote{Based on observations
  with the NASA/ESA \emph{Hubble Space Telescope} obtained at the
  Space Telescope Science Institute, which is operated by the
  Association of Universities for Research in Astronomy, Inc., under
  NASA contract NAS 5--26555.}  imaging data are available through a
broad red filter (F606W, F702W, and/or F814W) for 45 of the 59
clusters (Table~\ref{tab:sample}) of which 13 are drawn from new
LoCuSS ACS (PID:10881) and WFPC2 (PID:11312) observations.  The
reduction of the data on 10 clusters observed under PID:5701, PID:6488
and PID:8249 is described by \citet{smith05}.  Of the remaining 36
clusters, the 18 with WFPC2 data (PIDs:5352, 8301, 8719, 11312) were
all reduced onto a $0.1''$ pixel scale using {\sc wfixup, wmosaic,
  imshift} and {\sc crrej} tasks within {\sc iraf} to clean, register
and combine the individual exposures.  Details of the reduction of the
remaining clusters observed with ACS are described by Hamilton-Morris
et al.\ (2010, in preparation).

\subsection{\emph{Chandra} X-ray Observations}\label{sec:chandra}

\emph{Chandra} X-ray observations are available for 41 of the total
sample of 59 clusters.  The reduction and analysis of these data are
described in detail by \cite{sanderson09a}.  In brief, for each
cluster, an annular spectral profile was extracted and used to
deproject the X-ray emission to measure the gas density and
temperature in spherical shells.  The phenomenological cluster model
of \cite{ascasibar08} was then jointly fitted to the temperature and
density profiles to determine the mass profile, assuming hydrostatic
equilibrium, following the procedures described in
\cite{Sanderson10}. The model is based on a \cite{hernquist90}
density profile, which yields larger scale radii (and correspondingly
lower mass concentrations) than the commonly-used NFW profile.
Following \cite{sanderson09a}, we also use the logarithmic slope of
the gas density profile at 0.04 $r_{500}$
\citep[$\alpha$;][]{Vikhlinin07} as an indicator of cool core strength,
which has also been shown to correlate with the substructure fraction
of cluster cores, based on strong lens models \citep{richard10}.  A more
negative value of $\alpha$ indicates a steeper central gas density
profile, and thus a stronger cool core, and vice versa.

\subsection{Statistical Comparison of Sub-samples}\label{sec:tests}

Incomplete coverage of the parent sample of 115 clusters with WIRC,
and heterogeneous coverage of the WIRC-observed clusters with other
facilities (Fig.~\ref{fig:sample}) may introduce subtle biases into
our results.  We therefore compare statistically the various observed
sub-samples, including for completeness the sub-sample for which
\emph{Spitzer} data are available (\S\ref{sec:lumgap}).  Specifically,
the cluster X-ray luminosities are compared, after correction for the
modest redshift evolution within the sample due to the expansion of
the universe: $L_{X,z}=L_XE(z)^{-2.7}$, where
$E(z)=H(z)/H_0=[\Om(1+z)^3+\Ol]^{0.5}$ following \citet{Evrard02}.

The mean X-ray luminosity of the full sample of 59 clusters is
statistically indistinguishable from the mean luminosity of the
sub-samples observed at other wavelengths (Table~\ref{tab:stats}).  We
also draw 100,000 samples of 59 clusters at random from the combined
volume-limited samples defined by $0.15<z<0.2$,
$2.2\times10^{44}\ergs\le\Lxz\le4.2\times10^{44}\ergs$, and
$0.15<z<0.3$, $\Lxz\ge4.2\times10^{44}\ergs$ (see
Fig.~\ref{fig:sample}).  The average X-ray luminosity of the observed
sample of 59 clusters is well within one standard deviation of the
average X-ray luminosity of these randomized samples
(Table~\ref{tab:stats}).  We therefore conclude that the $L_X$
distributions of the full sample of 59 clusters, the sub-samples
observed with other telescopes, and the volume limited-sample defined
above, are all statistically indistinguishable from each other.  We
therefore expect any biases to be negligible, and that our results can
be treated as comparable with those that would be achieved with a
volume-limited sample.  We also take care to double check that the
$\dm$-distributions of the various observational sub-samples are
statistically indistinguishable from each other in
\S\ref{sec:analysis}.

\begin{table}
\begin{center}
\caption{Statistical Comparison of Sub-samples\label{tab:stats}}
\begin{tabular}{p{48.5mm}p{7.5mm}p{15mm}}
\hline
Sample & $N_{\rm clus}$ & $\mean{\log_{10}(L_X)}^a$ \\
\hline
All clusters observed with WIRC & $59$ & $44.71\pm0.03$ \\
Clusters observed with WIRC \& \emph{HST} & $45$ & $44.71\pm0.03$ \\
Clusters observed with WIRC \& \emph{Chandra} & $41$ & $44.76\pm0.04$ \\
Clusters observed with WIRC \& \emph{Spitzer} & $39$ & $44.73\pm0.03$ \\
Mean of 100,000 samples drawn randomly from volume-limited sample & $59$ & $44.71\pm0.03$ \\
\hline
\end{tabular}
\end{center} 
{\footnotesize $^a$~The uncertainties are errors on the
  mean X-ray luminosity of each sample, with the exception of the last
  row, in which we quote the standard deviation of the 100,000 samples
  around the mean luminosity of all of these randomly drawn samples.}
\end{table}

\section{Analysis and Results}\label{sec:analysis}

\subsection{Source Detection and Photometry}\label{sec:phot} 

The $J$- and $K$-band frames were analyzed with SExtractor
\citep{bertin96}, extracting all objects subtending ${>}25\pixel$ at
${\rm S/N}>2.5\pixel^{-1}$.  The resulting catalogs were matched using
a search radius comparable with the seeing disk, and point sources
were excluded based on the stellarity index calculated by SExtractor.
In the absence of spectroscopic redshift information we rely on the
red ridge line of galaxies seen in the $(J-K)/K$ colour-magnitude
diagrams for each cluster (e.g.\ Fig.~\ref{fig:cm}) to isolate likely
cluster galaxies.  A simple model based on redshifting local galaxy
spectral templates \citep{king85} confirms that the $(J-K)$
colour of galaxies varies by $\ls0.2\mags$ between E/S0 and
Scd spectral types.  We therefore selected galaxies within
$\pm0.2$\,magnitudes of the BCG colour in each cluster as likely
cluster members (see horizontal lines in Fig.~\ref{fig:cm}).

The extended envelope of the BCGs typically spans a diameter of
$\sim1{\rm arcmin}$ in the \emph{HST} frames.  In contrast, BCGs
typically span just $\sim20-30\arcsec$ in the near-infrared frames.
The difference is due to the brighter sky in the near-infrared
relative to the optical.  We therefore use the deep F702W
\emph{HST}/WFPC2 data available for 10 clusters in our sample
\citep{smith05} to estimate the $K$-band flux lost due to the bright
$K$-band sky, under the assumption that the $(R_{702}-K)$ colour of
BCGs does not vary significantly with radius on large scales.  This
assumption introduces negligible systematic uncertainty into our
results because colour gradients in elliptical galaxies are measured
to be $d(R-K)/d(\log r)\sim0.3-0.4$ \citep{labarbera04,labarbera10},
which translates into a possible $\sim0.1$~magnitude systematic error
on the factor of 2 radial corrections to the $K$-band photometry
estimated below.  

After masking out other galaxies from the data, the BCG $R_{702}$- and
$K$-band light distributions are modelled using {\sc ellipse} in the
{\sc stsdas} package in {\sc iraf}.  The $R_{702}$-band model is then
used to extrapolate the $K$-band light distribution out to $2\sigma$
above the mean local background. The same procedure was applied to a
sample of $L^\star$ galaxies detected in the WFPC2 frame of each of
these ten clusters.  This analysis revealed that reliance on solely
$K$-band data causes the the total flux of BCGs to be under-estimated
by $\sim0.3-0.7\mags$, with a median of $\sim0.45\mags$.  This effect
is much less severe for non-BCG's, with total flux being
under-estimated by $\sim0.07-0.15\mags$, with a median of $0.1\mags$.
We fit a straight-line to these data: $\Delta\,K=\alpha+\beta\,K$,
obtaining $\alpha=-1.67\pm0.43$ and $\beta=0.11\pm0.03$ .  The
correction, $\Delta\,K$, was then applied to all galaxies within our
sample.  The amplitude of this systematic correction to the luminosity
gap statistic measurements is therefore $\beta\dm$ and is typically in
the range $\sim0-0.3\,\mags$ with an uncertainty of $\sim25\%$, both
of which are smaller than the bin-width in our subsequent analysis.
Our results are therefore not significantly affected by the
uncertainties on this correction.

\subsection{Luminosity Gap Statistic of $10^{15}\Msol$ Clusters}
\label{sec:lumgap}

\begin{figure}
\centerline{
\epsfig{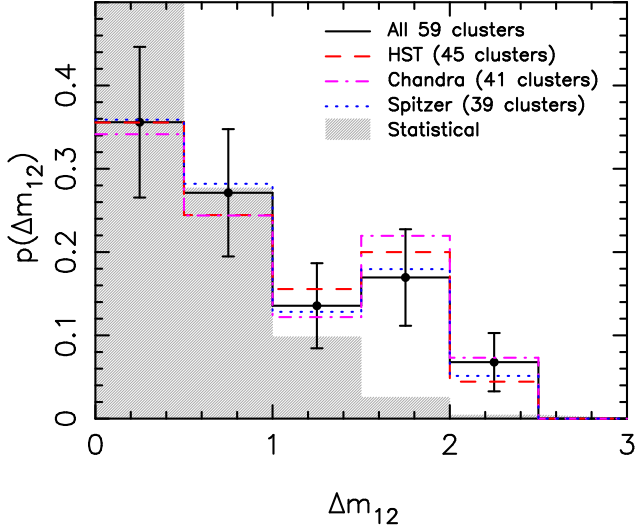}
}
\caption{Distribution of the observed luminosity gap (black points and
  solid line).  The gray filled histogram is the expected distribution
  if the galaxies are drawn at random from a Schechter function
  following \citet{dariush07} (see \S\ref{sec:lumgap} for more
  details).  The dashed, dot-dashed, and dotted histograms show the
  $\dm$-distributions of the sub-samples of clusters for which
  \emph{HST}, \emph{Chandra}, and \emph{Spitzer} data are available.}
\label{fig:dm12histobs}
\end{figure}

In the absence of models of the mass distribution, and thus
measurements of $r_{200}$ for all clusters in the sample we adopt a
fixed projected physical radius of $R=640\kpc$ within which to
calculate $\dm$ for each cluster.  This aperture fits comfortably
within the observed field of view for all clusters, and corresponds to
$\sim0.4r_{200}$ for a $M_{\rm virial}\simeq10^{15}\Msol$ cluster at
$z=0.2$.  The distribution of the luminosity gap statistic is shown in
Fig.~\ref{fig:dm12histobs}; $p(\dm)$ is a declining function of $\dm$.
We therefore fit a straight-line to the data: $p(\dm)=A+B\dm$,
weighting the data-points by $\sigma^{-2}$ where $\sigma$ is the
Poisson uncertainty on $\dm$ in each bin.  The best-fit parameter
values are: $A=0.36\pm0.03$ and $B=-0.13\pm0.02$.  We also measure the
fraction of ``fossil clusters'': a total of 4 clusters have $\dm\ge2$,
yielding a fraction of $10^{15}\Msol$ clusters satisfying this
selection of $p(\dm\ge2)=0.07^{+0.05}_{-0.03}$, where the error bar is
at $1\sigma$ using binomial statistics (Gehrels 1986).

Following \citet{dariush07} we also show in Fig.~\ref{fig:dm12histobs}
the $\dm$ distribution derived from a Monte Carlo simulation in which
galaxies were drawn at random from a Schechter function with
$M^\star=-24.5$ and $\alpha=-1.2$, adopted from a fit of the Schechter
function to the $K$-band galaxy luminosity function of the Millennium
semi-analytic catalogue, and is also consistent with observed
luminosity functions \citep[e.g.][]{lin04a}.  This simulation allows
us to identify whether the $\dm$ distribution presents any excess
probability over random statistical sampling of a common underlying
luminosity function.  Excess probability over random is only found at
$\dm\gs1$.  We measure the observed probability of a cluster to have a
luminosity gap of $\dm\ge1$ to be $p(\dm\ge1)=0.37\pm0.08$, compared
with the estimated probability based on the Monte Carlo simulation of
$p_{\rm MC}(\dm\ge1)=0.13$.  We therefore detect an excess probability
over random sampling at $\dm\ge1$ of $\sim0.24$ at $\sim3\sigma$
significance, and conclude that the $\dm$ distribution at $\dm>1$ has
a physical origin.

The $\dm$ distributions of the sub-samples of clusters for which
\emph{HST}, \emph{Chandra}, and \emph{Spitzer} data are available are
statistically consistent with that of the full sample of 59 clusters
(Fig.~\ref{fig:dm12histobs}).  Two sample Kolmogorov-Smirnov (KS)
tests that compare the \emph{HST}, \emph{Chandra}, and \emph{Spitzer}
sub-samples in turn with the full sample confirm that the probability
of the respective sub-samples being drawn from a different underlying
$\dm$ distribution than the full sample is $P\le0.1\%$ in all cases,
with the largest difference between the cumulative distributions being
$D=0.0748$, between the \emph{Chandra} sub-sample and the full sample.

\begin{figure}
\centerline{
  \epsfig{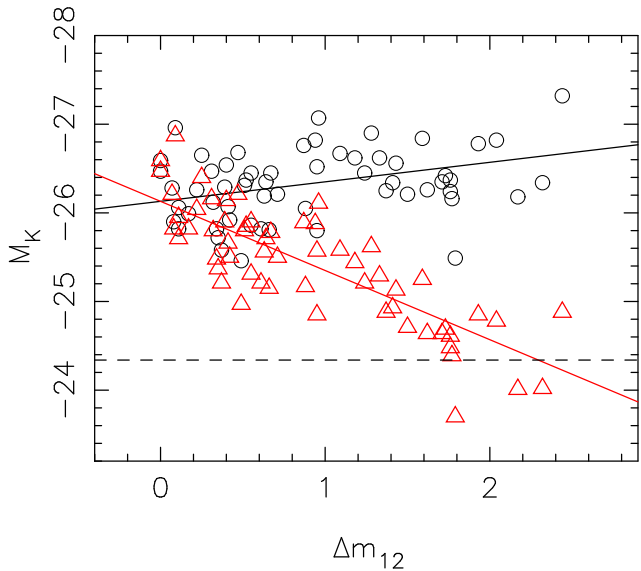}
}
\caption{Absolute $K$-band magnitude of the first (black circles) and
  second ranked (red triangles) galaxies as a function of luminosity
  gap.  The solid black and red lines show the best-fit straight-line
  to the data -- see \S\ref{sec:lumgap} for more details.  The
  horizontal dashed line is at $M_K=-24.34$, the absolute magnitude of
  an $L^\star$ galaxy, taken from \citet{lin04a}.}
\label{fig:m1dm12obs}
\end{figure}

We also look at how the absolute magnitude of the first and second
ranked cluster galaxies vary with $\dm$ (Fig.~\ref{fig:m1dm12obs}).
The luminosity of the first ranked galaxy increases very slowly with
$\dm$, remaining in the range $-27\ls M_K\ls-26$ across the full range
of $\dm$.  In contrast, the luminosity of the second ranked galaxy
declines from $M_K\sim-26$ at $\dm\sim0$ to $M_K\sim-24$ at
$\dm\sim2$.  We characterize these trends by fitting the following
relations to the data: $M_{K,1}=\alpha_1+\beta_1\dm$ and
$M_{K,2}=\alpha_2+\beta_2\dm$, where the numerical subscripts denote
the first and second ranked galaxies respectively.  The best-fit
values are: $\alpha_1=-26.13\pm0.02$, $\beta_1=-0.22\pm0.02$, and
$\alpha_2=-26.13\pm0.02$, $\beta_2=0.78\pm0.02$.  Empirically large
luminosity gap statistics are therefore due to both an over-bright
BCG, $M_{K,1}({\dm}{=}2){-}M_{K,1}({\dm}{=}0){\simeq}{-}0.4$, and an
under-bright second ranked galaxy,
$M_{K,2}({\dm}{=}2){-}M_{K,2}({\dm}{=}0){\simeq}1.6$.  The relative
faintness of second ranked galaxies in large luminosity gap clusters
supports the idea that the growth of dominant BCGs is driven by the
merging of luminous cluster galaxies with the BCG.  Indeed the current
SFR of BCGs discussed above lends additional support -- the BCG in a
cluster with a luminosity gap of $\dm=2$ is $6\times$ more luminous
and has a stellar mass of $\sim10^{11}\Msol$ more than the second
ranked galaxy.  Just two of the four clusters with $\dm\gs2$ in
Fig.~\ref{fig:dmalpha} host an active BCG.  The most active of these,
A\,1835, is forming stars at $\SFR=125\Msolpyr$ \citep{Egami06}, and
the other, RXJ\,2129.6$+$0005, is forming stars at $\SFR=14\Msolpyr$
\citep{Quillen08}.  These two BCGs would therefore have to form stars
continuously at this rate for $\sim10^9$ and $\sim10^{10}$\,years
respectively for their large luminosity gap to be caused exclusively
by gas cooling and consequent star formation.

Finally, we note that on average second ranked galaxies in clusters
with $\dm>2$ have $\mean{M_{K,2}}=-24.6\pm0.6$ where the uncertainty
is the rms scatter around the mean.  \cite{lin04a} measured
$M_K=-24.34\pm0.01$ for $L^\star$ cluster galaxies at $z\le0.1$, in
agreement with similar studies of field galaxies and of higher
redshift clusters \citep{depropris99,Cole01}.  The distribution of
luminosities of second ranked cluster galaxies in clusters with
$\dm>2$ is therefore statistically consistent with them being
$L^\star$ galaxies.  This contrasts with low mass $\dm>2$ systems,
i.e.\ fossil groups, in that $L^\star$ galaxies are absent from low
mass systems.  This difference is probably due, at least in part, to
the relative inefficiency of galaxy merging in massive clusters.

\subsection{Comparing Luminosity Gap with Cool Core
  Strength}\label{sec:coolcore}

\begin{figure}
\centerline{
\epsfig{file=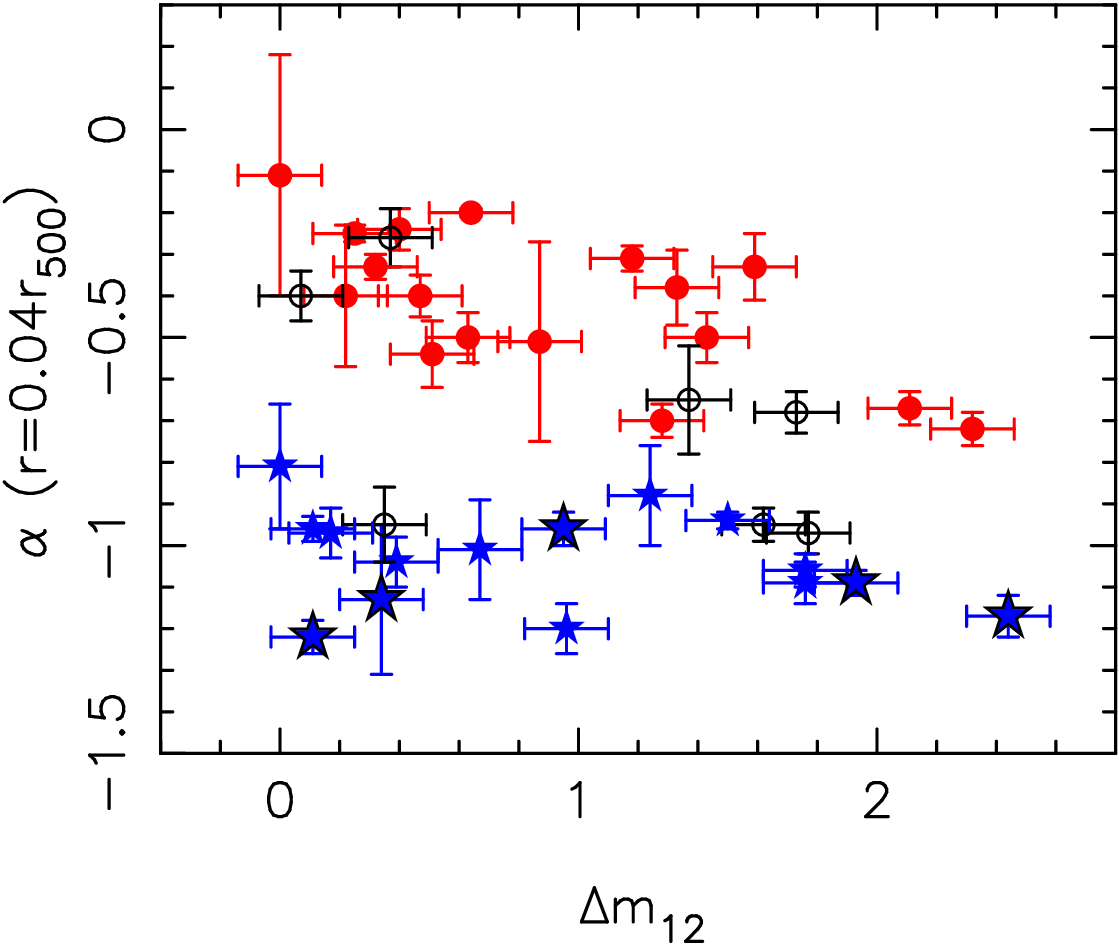,width=70mm,angle=-90}
}
\caption{The gradient of the logarithmic gas density profile at
  $0.04r_{500}$ versus luminosity gap for 41 clusters that have also
  been observed with \emph{Chandra} and studied by
  \citet{sanderson09a}.  Blue stars correspond to clusters with an
  H$\alpha$ emitting BCG \citep[see][]{sanderson09a}; blue stars with
  a black outline have also been identified as hosting a BCG that is
  forming stars at $\SFR\gs10\Msolpyr$ using \emph{Spiter}/MIPS
  observations; filled red circles denote clusters with BCGs that are
  not H$\alpha$ emitters and are forming stars at $\SFR<10\Msolpyr$;
  open black circles indicate clusters that have not been observed
  with \emph{Spitzer}.  }
\label{fig:dmalpha}
\end{figure}

To explore further the physical origin of large luminosity gaps we
plot $\dm$ versus $\alpha$, the slope of the logarithmic gas density
profile at $0.04r_{500}$, for 41 clusters that have also been observed
with \emph{Chandra} in Fig.~\ref{fig:dmalpha}.  The measurements of
$\alpha$ are based on \citeauthor{sanderson09a}'s
(\citeyear{sanderson09a}) analysis of the \emph{Chandra} data
(\S\ref{sec:chandra}).  At $\dm\simeq0$ the clusters span the full
range of cool core strengths: $-1.2\ls\alpha\ls-0.1$.  This dynamic
range shrinks to just $-1.2\ls\alpha\ls-0.6$ at $\dm\gs2$ -- the
clusters with large luminosity gaps also host relatively strong cool
cores.  We also identify star-forming BCGs in Fig.~\ref{fig:dmalpha}.
It has long been known that H$\alpha$ emission from the BCG is closely
associated with the presence of significant central cooling in the
cluster core \citep[e.g.][]{heckman81,crawford99}.  More recently,
\cite{sanderson09a} found in their sample of 65 clusters that
H$\alpha$ emitting BCGs occur exclusively in those clusters with the
most cuspy inner gas density profiles ($\alpha<-0.85$), and where the
projected offset between the X-ray centroid and the BCG is
$\le0.02r_{500}$.  The same is true of the five BCGs with star
formation rates (SFR) of $\gs10\Msolpyr$, based on mid-infrared
observations with \emph{Spitzer}/MIPS -- this SFR corresponds to a
flux of $\sim1\mJy$ from a BCG at $z\simeq0.2$.  These measurements
are drawn from the literature \citep{Egami06,Quillen08} and our own
measurements using data from Cycle~4 (PID:40827, PI: Smith; PID:
41011, PI: Egami) the details of which will be published elsewhere
(Egami et al., in prep.).  Fig.~\ref{fig:dmalpha} therefore confirms
that cool core clusters tend to host actively star-forming BCGs
\citep[e.g.][]{Edge99,Egami06,Quillen08}.  However, cool core clusters
($\alpha\ls-1$) with active BCGs (${\rm SFR}\gs10\Msolpyr$, and/or
H$\alpha$ emission) are found across the full range of $\dm$ in
Fig.~\ref{fig:dmalpha}.

\begin{figure*}
\centerline{
\epsfig{file=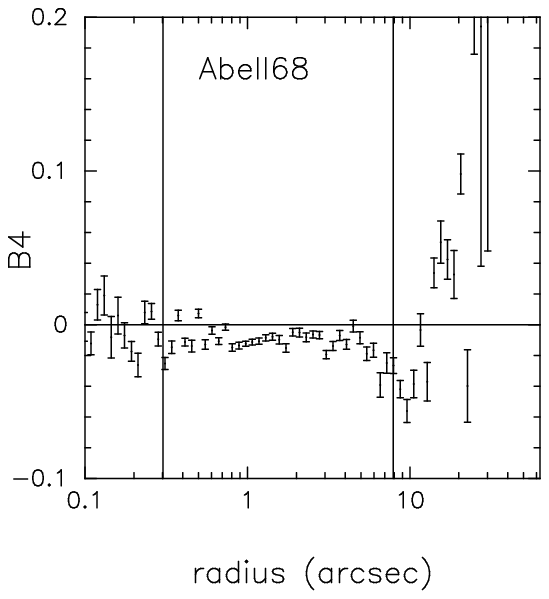,width=1.65in}
\epsfig{file=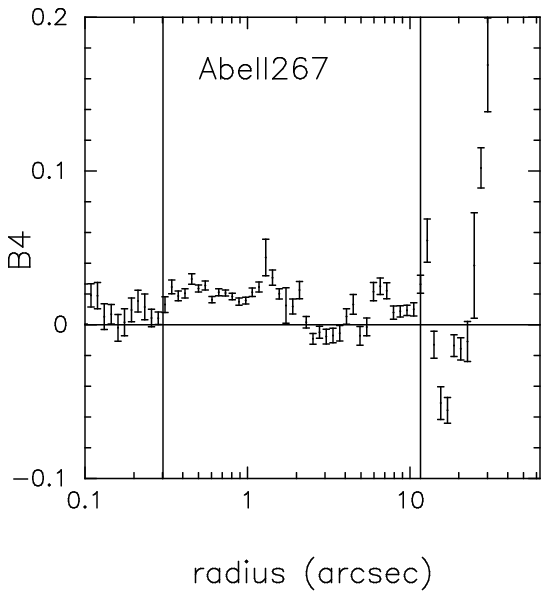,width=1.65in}
\epsfig{file=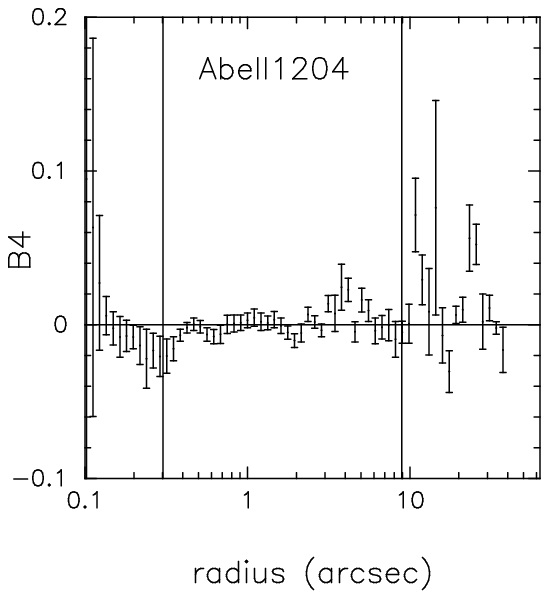,width=1.65in}
\epsfig{file=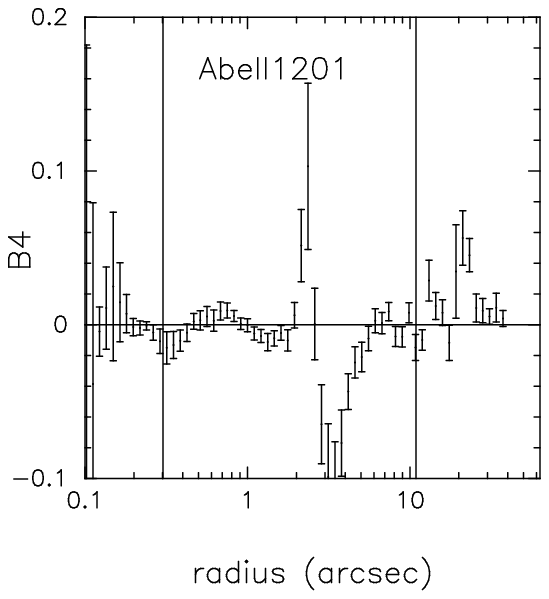,width=1.65in}
}
\caption{Example isophotal shape profiles. From left to right the BCGs
  are classified as boxy, disky, pure ellipse and unclassified. The
  vertical line at the left is set to $3\times$ the FWHM of point
  sources and the one at the right indicates the half-light radius of
  the BCG.  Note that to keep the analysis simple and conservative, no
  flux was masked out of the \emph{HST} data.  So, for example, the
  BCG in A\,1201 was unclassified because of the impact of the
  gravitational arc at a BCG-centric radius of $\sim2\arcsec$
  \citep{edge03} on the isophotal analysis. }
\label{fig:sampleiso}
\end{figure*}

These results are consistent with the interpretation of large
luminosity gap clusters as objects that formed relatively early, and
subsequently developed a large luminosity gap through the merging of
bright cluster galaxies with the BCG.  A similarly long period of time
-- a few Gyr -- is required to form a strong cool core following
cluster formation.  Conversely, if all clusters with smaller
luminosity gaps formed more recently than those with larger gaps, and
thus have had insufficient time to form a large luminosity gap and a
cool core, then they should all host relatively weak cool cores.
However this is not the case.  This can be understood if the so-called
``fossil'' status of a large luminosity gap cluster is not the
end-point of its evolution.  If bright ($L>L^\star$) galaxies fall
into a cool core ``fossil'' cluster, then that cluster would move
immediately leftward from the bottom right of Fig.~\ref{fig:dmalpha}.
As the in-falling system (presumably a group) reaches the cluster core
$\sim1\Gyr$ later, it may disrupt partially or fully the cooling of
gas onto the BCG, and cause the cluster to move vertically in the
$\dm-\alpha$ plane.  This scenario naturally explains the triangular
distribution of points in Fig.~\ref{fig:dmalpha}, and is consistent
with hierarchical infall (i.e.\ mergers) playing a role in regulating
cooling in cluster cores.

To place this discussion on a more quantitative footing we adopt a
strategy that we return to often in \S\ref{sec:analysis} -- we split
the sample into low- ($\dm<1$) and high-$\dm$ ($\dm>1$) sub-samples
and perform a two sample KS test on the cumulative distribution of the
other variable, in this case $\alpha$.  The hypothesis that high-$\dm$
clusters are drawn from the same underlying $\alpha$ distribution as
low-$\dm$ clusters is rejected at just $74\%$ confidence, i.e.\
slightly over $1\sigma$ significance, based on a maximum difference
between the cumulative $\alpha$ distributions of $D=0.3102$.  In the
absence of a decisive test, we therefore divide the sample at
$\dm=1.5$, i.e.\ a more extreme value of $\dm$, attempting to identify
roughly the luminosity gap at which the $\alpha$ distribution diverges
from that of lower-$\dm$ clusters.  This time the two sample KS test
rejects the null hypothesis at $95\%$ confidence -- i.e.\ $2\sigma$ --
based on a maximum difference between the respective cumulative
$\alpha$ distributions of $D=0.472$.

\subsection{ Comparing the Luminosity Gap with BCG Morphology
}\label{sec:isophot}

\begin{table*}
\begin{center}
\caption{Results from Isophotal Analysis of Brightest Cluster Galaxies
  using \emph{HST} data. 
\label{tab:hstsample}
}
\begin{tabular}{lrlrllrl}
\hline
Cluster               & Effective      & ~~~ & \multispan2{\dotfill Extremum Method\dotfill}          & ~~~ & \multispan2{\dotfill Mean Method\dotfill} \\
                      & radius         & & $100B_{\rm 4,ext}$ & Classification          & & $100\mean{B_4}$         & Classification \\
                      & (arcsec)       & \\
\hline
A\,68                 & $7.9$  & & ${-}2.0{\pm}0.4$   & Boxy           & & ${-}0.99{\pm}0.19$ & Boxy       \\
A\,115                & $4.1$  & & ...                & Unclassified   & & ${+}0.96{\pm}0.22$ & Disky      \\
A\,141                & $3.6$  & & ${+}3.4{\pm}0.9$   & Disky          & & ${+}1.09{\pm}0.19$ & Disky      \\
ZwCl\,0104.4$+$0048   & $4.1$  & & ...                & Unclassified   & & ${-}0.14{\pm}0.80$ & Elliptical   \\
A\,209                & $7.4$  & & ${+}1.8{\pm}0.5$   & Disky          & & ${+}0.45{\pm}0.07$ & Disky   \\
A\,267                & $11.6$ & & ${+}3.5{\pm}1.2$   & Disky          & & ${+}0.99{\pm}0.17$ & Disky  \\
A\,291                & $7.3$  & & ${+}2.1{\pm}0.6$   & Disky          & & ${+}0.67{\pm}0.24$ & Disky  \\
A\,383                & $8.3$  & & ...                & Unclassified   & & ${-}0.20{\pm}0.97$ & Unclassified \\
RXC\,J0331.1$-$2100   & $3.1$  & & ...                & Unclassified   & & ${-}2.69{\pm}7.50$ & Unclassified \\
A\,521                & $3.9$  & & ${-}3.1{\pm}1.0$   & Boxy           & & ${-}0.44{\pm}0.79$ & Elliptical\\
A\,586                & $7.7$  & & ...                & Unclassified   & & ${-}0.16{\pm}0.64$ & Elliptical \\
ZwCl\,0740$+$1740     & $5.5$  & & ${+}0.1{\pm}0.2$   & Elliptical     & & ${-}0.11{\pm}0.06$ & Boxy \\
A\,611                & $3.5$  & & ...                & Unclassified   & & ${-}0.21{\pm}0.15$ & Boxy \\
ZwCl\,0839.9$+$2937   & $3.5$  & & ...                & Unclassified   & & ${+}0.84{\pm}0.50$ & Disky\\
ZwCl\,0857.9$+$2107   & $4.6$  & & ${-}4.0{\pm}1.1$   & Boxy           & & ${-}1.49{\pm}0.61$ & Boxy  \\
A\,750                & $4.0$  & & ${+}5.0{\pm}1.5$   & Disky          & & ${+}0.11{\pm}1.45$ & Unclassified \\
A\,773                & $6.8$  & & ${-}2.1{\pm}0.4$   & Boxy           & & ${-}0.50{\pm}0.18$ & Boxy  \\
ZwCl\,0949.6$+$5207   & $7.1$  & & ${+}2.4{\pm}0.7$   & Disky          & & ${+}0.22{\pm}0.30$ & Elliptical \\
A\,901                & $3.2$  & & ${+}1.1{\pm}0.3$   & Disky          & & ${+}0.50{\pm}0.10$ & Disky \\
RX\,J1000.5$+$4409    & $3.1$  & & ...                & Unclassified   & & ${-}0.77{\pm}0.18$ & Boxy \\
A\,963                & $14.3$ & & ...                & Unclassified   & & ${+}0.46{\pm}0.19$ & Disky\\
A\,1201               & $10.8$ & & ...                & Unclassified   & & ${-}0.26{\pm}0.48$ & Elliptical \\
A\,1204               & $8.9$  & & $0.0{\pm}0.7$      & Elliptical     & & ${-}0.10{\pm}0.21$ & Elliptical \\
A\,1246               & $8.6$  & & ...                & Unclassified   & & ${-}0.14{\pm}1.78$ & Unclassified \\
A\,1423               & $7.4$  & & ...                & Unclassified   & & ${-}0.91{\pm}60.6$ & Unclassified  \\
ZwCl\,1231.4$+$1007   & $4.4$  & & $-1.3{\pm}0.5$     & Boxy           & & ${-}0.56{\pm}0.19$ & Boxy \\
A\,1682               & $6.2$  & & $0.0{\pm}0.5$      & Elliptical     & & ${-}0.43{\pm}0.27$ & Boxy  \\
A\,1763               & $8.2$  & & $0.0{\pm}0.6$      & Elliptical     & & ${+}0.04{\pm}0.08$ & Elliptical \\
A\,1835               & $6.8$  & & ...                & Unclassified   & & ${+}1.10{\pm}0.60$ & Disky \\
A\,1914               & $12.2$ & & $0.0{\pm}0.5$      & Elliptical     & & ${+}0.25{\pm}0.15$ & Disky \\
MS\,1455.0$+$2232     & $5.0$  & & ...                & Unclassified   & & ${-}0.47{\pm}0.25$ & Boxy  \\
A\,2009               & $9.3$  & & ...                & Unclassified   & & ${+}0.14{\pm}0.19$ & Elliptical \\
A\,2146               & $6.3$  & & ...                & Unclassified   & & ${-}0.20{\pm}0.50$ & Elliptical \\
A\,2204               & $7.9$  & & ...                & Unclassified   & & ${+}0.08{\pm}1.20$ & Unclassified  \\
A\,2218               & $8.1$  & & ${+}2.3{\pm}0.8$   & Disky          & & ${+}0.83{\pm}0.29$ & Disky       \\
A\,2219               & $8.5$  & & ...                & Unclassified   & & ${+}0.83{\pm}0.19$ & Disky        \\ 
A\,2254               & $9.6$  & & ...                & Unclassified   & & ${-}0.03{\pm}0.36$ & Elliptical \\
RXJ\,1720.1$-$2638    & $5.5$  & & ...                & Unclassified   & & ${+}0.28{\pm}1.39$ & Unclassified \\
A\,2261               & $7.2$  & & ...                & Unclassified   & & ${+}0.39{\pm}7.20$ & Unclassified \\
A\,2345               & $7.8$  & & $0.0{\pm}0.5$      & Elliptical     & & ${+}0.03{\pm}0.09$ & Elliptical   \\
RX\,J2129.6$+$0005    & $9.6$  & & ...                & Unclassified   & & ${-}0.16{\pm}0.60$ & Elliptical \\
A\,2390               & $3.5$  & & ${+}3.5{\pm}0.5$   & Disky          & & ${+}1.42{\pm}0.32$ & Disky      \\
A\,2485               & $3.9$  & & ${-}0.5{\pm}0.6$   & Elliptical     & & ${+}0.19{\pm}0.63$ & Elliptical      \\
A\,2537               & $5.2$  & & $0.0{\pm}0.3$      & Elliptical     & & ${-}0.11{\pm}0.18$ & Elliptical \\
A\,2631               & $6.0$  & & $0.0{\pm}0.5$      & Elliptical     & & ${-}0.24{\pm}0.12$ & Boxy \\
\hline
\end{tabular}
\end{center}
\end{table*}

We use the high angular resolution \emph{HST} imaging observations of
the 45 clusters discussed in \S\ref{sec:hst} to measure the isophotal
shape of the BCGs in these clusters.  The {\sc ellipse} task in {\sc
  IRAF} was used to measure the fourth Fourier coefficient ($B_4$) of
the light distribution.  This coefficient indicates whether the galaxy
has a disky or boxy shape \citep{bender88}.  In
Fig.~\ref{fig:sampleiso} we show the $B_4$ profile of four BCGs to
illustrate the diversity within the sample.  Following \cite{bender89}
we tried to use the extremum value of $B_4$ (i.e.\ $B_{\rm 4,ext}$ in
Table~\ref{tab:hstsample}) to classify galaxies as either disky
($B_{\rm 4,ext}>0$) or boxy ($B_{\rm 4,ext}<0$).  If the $B_4$ profile
passes through a stationary point, then the extremum is obtained by
finding the maximum or minimum value of $B_4$ in the radial range
enclosed by $3\times$ the FWHM of point sources and the effective
radius derived from a de~Vaucouleurs profile fit.  In the absence of a
stationary point, the extremum value of $B_4$ is the value at the
effective radius, under the assumption that $B_4$ is a monotonic
function of radius.  However $B_4$ is in general not a monotonic
function of radius for BCGs in our sample, even for those with
isophotes that have, on average, boxy and disky isophotes
(Fig.~\ref{fig:sampleiso}).  For these reasons, the isophotal shapes
of 22 out of 45 BCGs cannot be classified based on $B_{\rm 4,ext}$.
We also find some clusters (e.g.\ A\,1204 -- see
Fig.~\ref{fig:sampleiso}) in which $B_4$ is consistent with zero
across the full radial range of the data.

We therefore implement a modified scheme, in which we calculate the
error-weighted mean value of $B_4$ in the same radial range as above,
with no weighting of the bins to account for the variation of the bin
solid angle as a function of radius.  BCGs with $\mean{B_4}$
consistent with zero within the uncertainties were classified as
elliptical, otherwise BCGs are classified as Boxy or Disky if
$\mean{B_4}{<}0$ or $\mean{B_4}{>}0$ respectively.  Finally, a BCG is
``Unclassified'' if the error on $\mean{B_4}$ is comparable with the
dynamic range of the data, i.e.\ ${\ge}0.01$.  BCG morphologies
derived under both Bender et al.'s ``extremum'' scheme and our own
``mean'' scheme are listed in Table~\ref{tab:hstsample} along with the
Boxy/Disky/Elliptical/Unclassified classification based on each
method.  The respective methods agree on morphological classification
for 16 of the 22 BCGs for which classification was possible under both
methods.  However, only three of the six discrepant BCGs have
$\mean{B_4}$ and $B_{\rm 4,ext}$ values that formally disagree between
the methods within the quoted uncertainties -- A\,521, A\,750 and
ZwCl\,0949.6$+$5207.  The important advantage of our method is that
classification is possible for an additional 15 BCGs that were
unclassifiable under the Bender et al.\ scheme.  We therefore adopt
$\mean{B_4}$ as our measure of BCG morphology for all clusters with
\emph{HST} data for the reasons outlined above regarding the general
absence of clearly defined stationary points and monotonic behaviour
of the $B_4$ profiles.

\begin{figure}
\centerline{
\epsfig{file=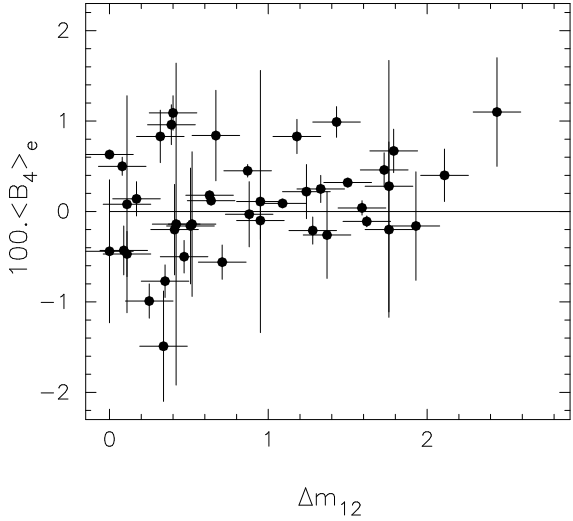,width=75mm}
}
\caption{Luminosity gap statistic (${\dm}$) versus error-weighted mean
  fourth Fourier component of the BCG light distribution
  ($\mean{B_4}$).  Positive values of $\mean{B_4}$ correspond to Disky
  BCGs; negative values correspond to Boxy BCGs; values consistent
  with zero are consistent with elliptical isophotes.  Clusters with
  $\dm\ls1$ host BCGs with both Boxy and Disky isophotes.  In contrast
  clusters with $\dm\gs1$ host only non-Boxy (i.e.\ Elliptical or
  Disky BCGs).}
\label{fig:isophot}
\end{figure}

In summary, out of 45 clusters, 10 are classified as Boxy, 13 as
Disky, 14 as Elliptical, and 8 are Unclassified.  In Fig
\ref{fig:isophot} we plot $\mean{B_4}$ versus ${\dm}$, the most
striking feature of which is the lack of clusters with large $\dm$ and
negative $\mean{B_4}$, i.e.\ boxy BCGs appear not to live in large
luminosity gap clusters.  As in \S\ref{sec:coolcore}, we split the
clusters into low-$\dm$ ($\dm\le1$) and high-$\dm$ ($\dm>1$) samples
and perform a two-sample KS test.  The low- and high-$\dm$ samples
contain $27$ and $18$ clusters respectively, with a maximum difference
between their cumulative $\mean{B4}$-distributions of $D=0.3567$.  The
hypothesis that the low- and high-$\dm$ samples are drawn from the
same underlying $\mean{B_4}$ distribution is therefore disfavoured at
$91\%$ confidence, i.e.\ $1.7\sigma$.  Unlike the situation for the
analysis of the $\alpha$ distributions of high- and low-$\dm$ clusters
in \S\ref{sec:coolcore}, the significance with which the null
hypothesis is rejected does not increase if the sub-samples are
re-defined by splitting the full sample at $\dm=1.5$.  This is obvious
from a comparison of Figs.~\ref{fig:dmalpha}~\&~\ref{fig:isophot}, and
suggests that the relationship between $\dm$ and BCG morphology is
stronger than between $\dm$ and cool core strength.

\subsection{Comparing the Luminosity Gap with Cluster Concentration
}\label{sec:conc}

\begin{figure}
\centerline{
\epsfig{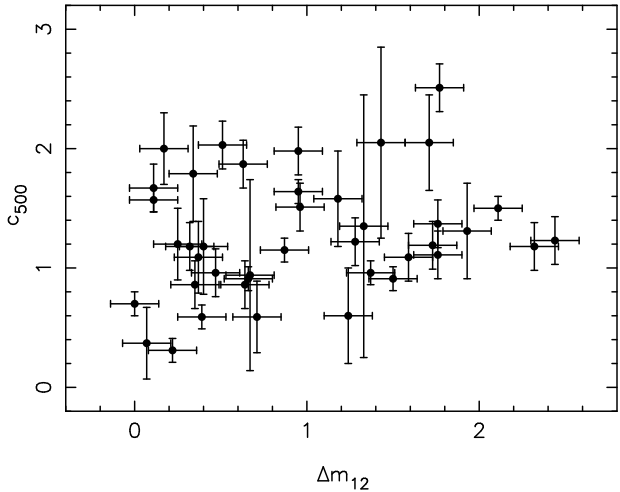}
}
\caption{Concentration $c_{500}$ versus luminosity gap for 41 clusters
  for which the X-ray-based mass profiles are available from
  \citeauthor{sanderson09a}'s (\citeyear{sanderson09a}) analysis of
  archival \emph{Chandra} data.  }
\label{fig:conc}
\end{figure}

To investigate the possibility that high-$\dm$ clusters formed at
earlier times than low-$\dm$ clusters, we explore the relationship
between $\dm$ and the shape of the cluster dark matter halos via the
concentration parameter.  In Fig.~\ref{fig:conc} we plot $\dm$ versus
concentration, $c_{500}$ for the 41 clusters with available
\emph{Chandra} data \cite{sanderson09a} (\S\ref{sec:chandra}).  The
$c_{500}-\dm$ distribution is similar to the $\mean{B4}-\dm$
distribution in that the lower-right of both plots is empty, and that
clusters with $\dm<1$ span the full dynamic range in the vertical
axis.  To quantify this we again perform a two sample KS test, on the
$\dm<1$ and $\dm>1$ sub-samples.  In this case the low- and high-$\dm$
samples contain $24$ and $17$ clusters respectively, with a maximum
difference between their cumulative $c_{500}$-distributions of
$D=0.2574$.  Acceptance/rejection of the null hypothesis that low- and
high-$\dm$ clusters are drawn from same underlying
$c_{500}$-distribution therefore have roughly equal probability.
However if we modify the definition of the low- and high-$\dm$
sub-samples by splitting the full sample at $\dm=1.5$ we are able to
reject the null hypothesis at $\sim1.7\sigma$.  We therefore conclude
that the $c_{500}-\dm$ plane qualitatively supports the interpretation
of the $\mean{B4}-\dm$ plane, however statistically this is not
decisive.  Specifically, clusters with a large luminosity gap tend to
have a relatively large concentration parameter, although there is a
curious deficit of clusters with $\dm\gs1.8$ and $c_{500}\gs1.5$.
Clusters with lower luminosity gaps plausibly comprise both clusters
that formed more recently than clusters with large gaps -- and thus
have lower concentration parameters -- and clusters that used to have
a large luminosity gap, and thus formed early, and have a higher
concentration parameter, but that then suffered infall of bright
($L>L^\star$) galaxies.  Put another way, the existence of clusters in
the top left corner of Fig.~\ref{fig:conc} is consistent with the
timescale on which the concentration parameter of a cluster may be
reset following a cluster-cluster merger being long compared with the
infall timescale of $\sim1\Gyr$.

\subsection{Comparing the Luminosity Gap with Cluster
  Substructure}\label{sec:substr}

\begin{figure}
\centerline{
\epsfig{file=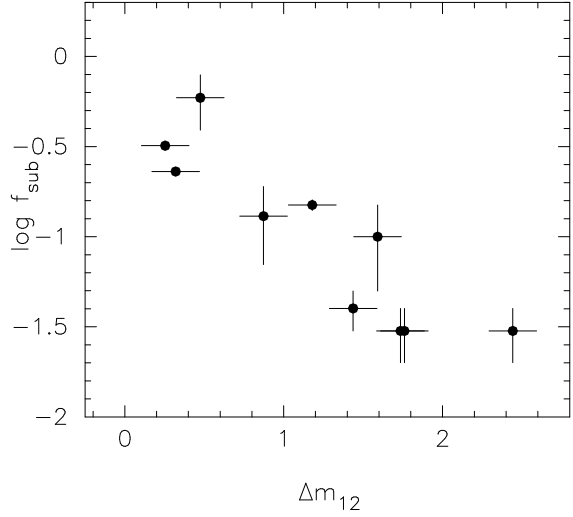,width=75mm}
}
\caption{Luminosity gap versus substructure fraction measured within
  $R\le250h^{-1}\kpc$ in 10 clusters from our sample by
  \citet{smith05}.}
\label{fig:fsub}
\end{figure}

Measurements of the substructure fraction ($\fsub$), i.e.\ the
fraction of the total cluster mass that resides in substructures, are
available for ten of the clusters \citep{smith05} from our sample of
59.  \citeauthor{smith05}'s gravitational lens models include mass
components that account explicitly for substructures required to
reproduce the observed positions of multiply-imaged background
galaxies -- these substructures comprise both galaxy group and
individual galaxy masses.  We plot $\fsub$ versus $\dm$ for these ten
clusters in Fig.~\ref{fig:fsub}, revealing a relationship between
these quantities in the sense that clusters with simpler gravitational
potentials (low ${\fsub}$) have more dominant BCGs (high $\dm$), and
vice versa.  To quantify this relationship, we fit a simple model to
the data: $\log\fsub=\mu+\nu\dm$, and obtain best-fit parameters of:
$\mu=-0.29\pm0.15$ and $\nu=-0.58\pm0.11$.  This result is consistent
with that found by \cite{richard10}, despite the smaller aperture of
$250\kpc$ used in their study.  This consistency arises because the
typical projected separation of the first and second ranked galaxies
in our sample is $\ls250\kpc$.  We also double-check that the $\dm$
distribution of the 10 clusters in Fig.~\ref{fig:fsub} is consistent
with that of the full sample, finding a maximum difference between the
cumulative $\dm$ distributions of $D=0.2414$, indicating roughly equal
probability of rejection/acceptance of the hypothesis that the two
samples are drawn from different underlying populations.  

\subsection{Summary}\label{sec:summary}

We now summarize the comparison of our luminosity gap measurements
with other probes of the structure, and thus the age and assembly
history of clusters, and discuss the interpretation of these results.

The clearest empirical relationship found is between the $\dm$ and
$\fsub$ in the sense that clusters with a dominant BCG ($\dm>1$) have
a lower substructure fraction ($\fsub<0.1$) and vice versa.  The
strong correlation between $\dm$ and $\fsub$ is in stark contrast with
the triangular distributions of clusters in the $\dm$-$\alpha$,
$\dm$-$\mean{B_4}$, and $\dm$-$c_{500}$ planes.
A simple physical interpretation of the $\dm$-$\fsub$ relation is that
both quantities are sensitive to the same thing.  As galaxies and
groups of galaxies fall into clusters the light emitted by the
galaxies will either cause $\dm$ to decrease or stay the same,
depending on how bright the infalling galaxies are.  At the same time
the total mass of these galaxies and the group-scale halos within
which they may be embedded causes $\fsub$ to increase.  Early studies
discussed the idea that galaxy groups with $\dm\ge2$ may have formed
at earlier times than groups with $\dm<2$.  However, more recently, a
variety of studies have shown that both $\dm$ and $\fsub$ are
correlated with \emph{both} the formation epoch of the host dark
matter halo, \emph{and} the recent hierarchical assembly history of
the halo \citep{dariush07,dariush10,smith08}.  Therefore both
theoretical and observational studies across a broad range of dark
matter halo mass are converging on the view that ``fossil'' status is
not an end-point in the evolution of galaxy systems that formed early.
Rather it is a phase that a galaxy system can evolve through if it
formed early and then suffered minimal hierarchical infall after the
formation of a bright massive central galaxy.  The triangular
distribution of clusters in the $\dm$-$\alpha$, $\dm$-$\mean{B_4}$,
and $\dm-c_{500}$ planes are all consistent with this interpretation,
and inconsistent with the idea that fossil galaxy systems are
evolutionary cul-de-sacs.  Specifically, if a cluster forms early and
then sufficient time elapses for a large luminosity gap to form via
merging of $L^\star$ gas-rich galaxies to form the BCG, and for a cool
core to form, then this cluster will reside in the top-right corner of
Figs.~7~\&~8 and the bottom right of Fig.~5.  If a $\gs L^\star$
galaxy then falls into the cluster, either on its own or in a group,
then the cluster would move left-ward in all of Figs~5, 7, and 8 as
soon as the infalling galaxy system crosses the aperture within which
$\dm$ is measured (in our case $0.4r_{200}$).  Several Gyr later the
infalling structure will reach the center of the cluster, and its
merger with the cluster may be sufficiently energetic to modify the
strength of the cluster cool core, the shape of the BCG, and the
concentration of the cluster dark matter halo.  In this way, clusters
can move vertically in Figs.~5, 7, and 8, and produce the observed
triangular distribution of clusters.

\begin{figure}
\centerline{
\epsfig{file=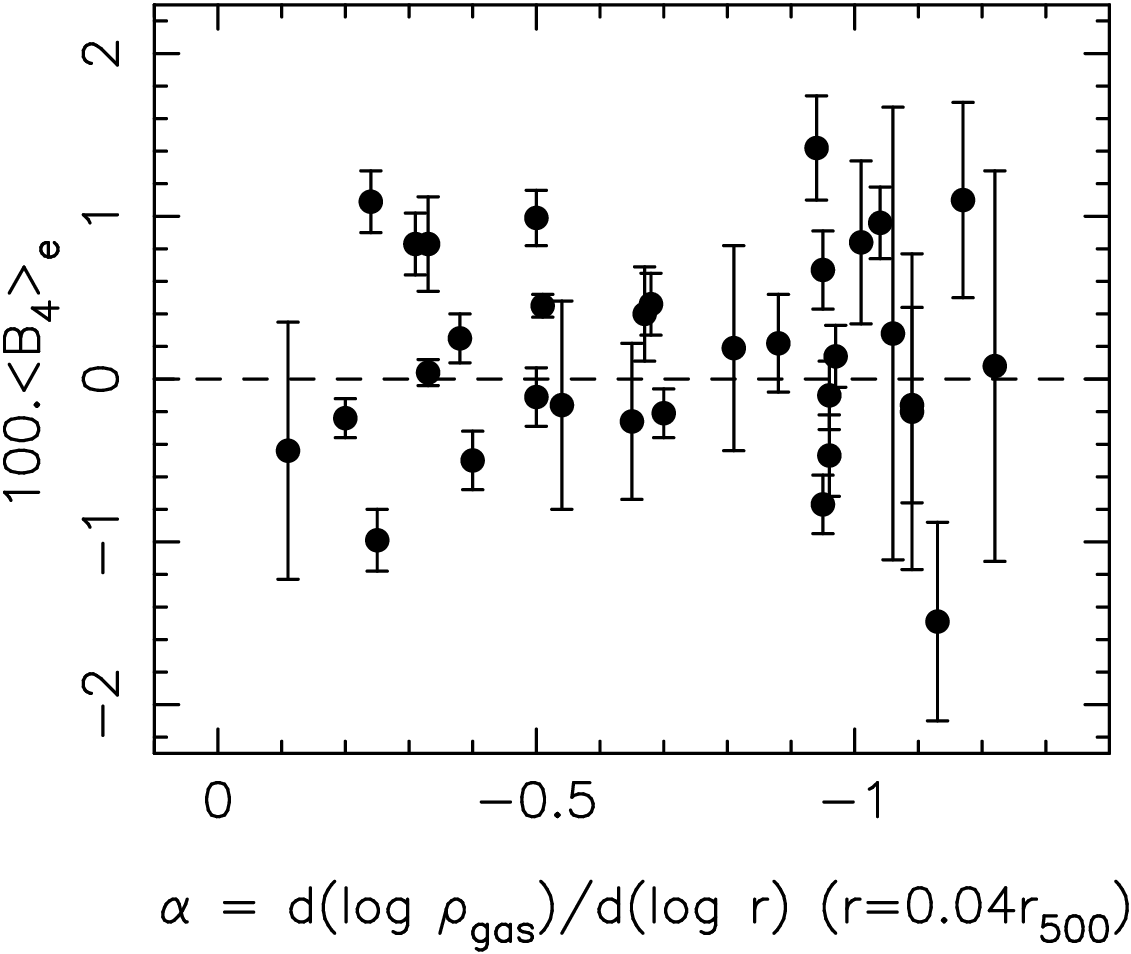,width=75mm,angle=-90}
}
\caption{Strength of the cool core in each cluster, as measured by
  $\alpha$ the slope of the logarithmic gas density profile at
  $0.04{\rm r_{500}}$ from \citet{sanderson09a} versus the
  $\mean{B_4}$.  The absence of a relationship between $\alpha$ and
  $\mean{B_4}$ suggests that disky BCG isophotes are more likely
  caused by such BCGs being formed from mergers between gas rich
  galaxies than by cooling of gas onto the BCG.  The typical error bar
  on $\alpha$ is $\ls0.1$.}
\label{fig:b4alpha}
\end{figure}

The interpretation of non-boxy morphologies ($\mean{B_4}{\ge}0$) of
BCGs in clusters with large luminosity gaps is an important element of
the discussion above.  \citet{khochfar05} showed that the morphology
of early-type galaxies is sensitive to the morphology (indicative of
gas content) of their progenitors and subsequent gas infall.  The
straightforward interpretation of the observables is therefore that
dominant BCGs formed from mergers of gas rich (presumably spiral)
galaxies and/or have accreted gas since the last major merger in their
assembly history.  Formation of dominant BCGs from gas rich
progenitors is consistent with the early formation of these BCGs as
discussed above, because at earlier times the galaxies from which BCGs
formed would have been more gas rich than at later times.

To disentangle the relative contribution of gas rich mergers and
accretion of gas to the disky shape of some BCGs we plot in
Fig.~\ref{fig:b4alpha} $\alpha$, the slope of the logarithmic gas
density profile at $0.04{\rm r_{500}}$ versus $\mean{B_4}$.  If BCG
morphology is strongly influenced by gas cooling onto the BCG then one
would expect a relationship between $\alpha$ and $\mean{B_4}$ in the
sense that disky BCGs ($\mean{B_4}>1$) would live in clusters with a
steep central ($\alpha<-0.5$) gas density profile.  This is because
clusters with steep central gas density profiles host a cool core --
i.e.\ a central positive temperature gradient, absence of an entropy
core, and a cooling timescale short compared with the age of the
universe.  However we do not find any strong relationship between
$\mean{B_4}$ and $\alpha$ in Fig.~\ref{fig:b4alpha}.  We divide the 41
clusters with \emph{Chandra} data into those with the strongest cool
cores -- $\alpha<-0.9$ -- and the rest.  A two sample KS test on these
two sub-samples yields a maximum difference between the cumulative
$\mean{B_4}$-distributions of $D=0.209$, indicating roughly equal
probability of accepting/rejecting the hypothesis that the two
distributions are drawn from the same underlying distributions.

In the absence of a strong relationship between cool core strength and
BCG morphology, we therefore conclude that BCG morphology is more
sensitive to the gas content of the galaxies that merged to form it,
than to the subsequent gas accretion history of the BCG.  This view is
consistent with the comparison of $\alpha$, $\dm$ and BCG activity in
Fig.~\ref{fig:dmalpha} and the discussion of the dependence of
$M_{K,2}$ on $\dm$ in \S\ref{sec:lumgap}.  The key point being that
the merging of luminous cluster galaxies to form the BCG appears to
have a much stronger influence on luminosity gap than gas cooling and
subsequent star formation within BCGs.

\section{Comparison with Theoretical Predictions}
\label{sec:theory}

Modern galaxy formation and evolution models contain physical
prescriptions for many physical processes relevant to the formation
and evolution of galaxies, including dynamical friction, conversion of
cold gas into stars during galaxy mergers, and AGN feedback.  These
processes are particularly important in the centres of galaxy clusters
where they regulate the cooling of gas onto the most massive galaxies
in the universe -- BCGs.  However the models were not constrained by
the luminosity gap distribution; our observational results can
therefore provide a strong test of the models.

\begin{figure}
  \centerline{
    \epsfig{file=f11.ps,width=75mm,angle=-90}
  }
  \caption{Distribution of the observed luminosity gap (black points --
    see also Fig.~\ref{fig:dm12histobs}) compared with the same for
    clusters with $\Mvir\ge5\times10^{14}\Msol$, measured within a
    projected BCG-centric radius of $640\kpc$ using the the Millennium
    simulation-based semi-analytic galaxy formation models of
    \citet{bower06}, \citet{croton06}, and \citet{delucia07}. The error bar
    on each bin in the theoretical histograms is comparable with the
    observational errors.}
  \label{fig:dm12histsim}
\end{figure}

\begin{figure*}
\centerline{
  \epsfig{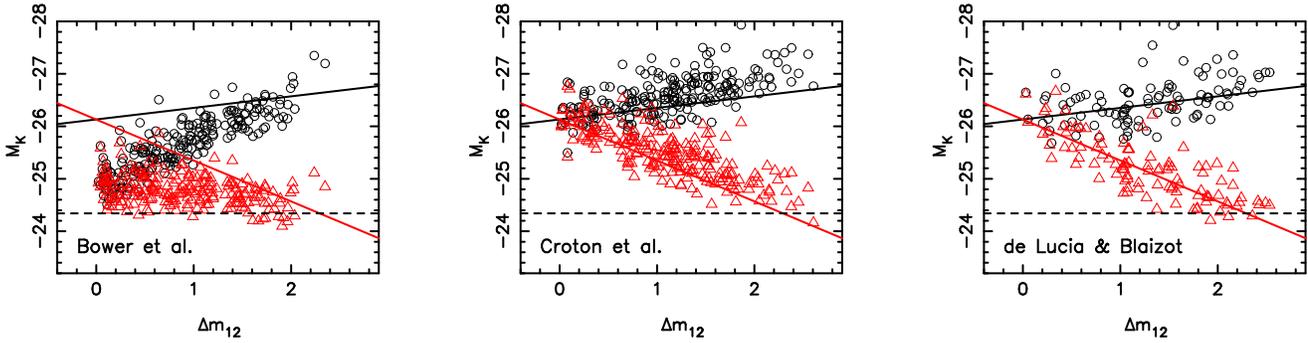}
}
\caption{Absolute $K$-band magnitude of the first (black circles) and
  second ranked (red triangles) galaxies as a function of luminosity
  gap from the three semi-analytic galaxy formation models discussed
  in \S\ref{sec:theory}.  The solid black and red lines show the best
  straight-line fit to the observational data shown in
  Fig.~\ref{fig:m1dm12obs}.  The horizontal dashed line in each panel
  is at $M_K=-24.34$, the absolute magnitude of an $L^\star$ galaxy,
  taken from \citet{lin04a}.}
\label{fig:m1dm12sim}
\end{figure*}

We compare our observations with the \cite{bower06}, \cite{croton06},
and \cite{delucia07} semi-analytic models, all of which are based on
the Millennium Simulation\footnote{The Millennium Simulation used in
  this paper was carried out by the Virgo Supercomputing Consortium at
  the Computing Centre of the Max-Planck Society in Garching. The
  semi-analytic galaxy catalogue is publicly available at
  http://www.mpa-garching.mpg.de/galform/agnpaper.} -- a cosmological
numerical simulation of dark matter in a volume spanning $500\hMpc$
containing $\sim10^{10}$ particles.  An important difference between
the models is that the \citeauthor{bower06} model implements
``quasar'' mode AGN feedback, whereas the \citeauthor{croton06} and
\citeauthor{delucia07} models implement ``radio'' mode AGN feedback.
We also note that \citeauthor{delucia07} compared their model
predictions with the observed properties of BCGs, however they didn't
compare with observed luminosity gaps.

First we select dark matter halos from the Millennium dark matter
friends of friends catalogue. Within the whole simulated volume, 209
hales were found with masses greater than $5\times10^{14}\Msol$,
i.e.\ above the mass threshold of the observed sample.  We then
extracted galaxies in these 209 halos from the semi-analytic galaxy
catalogues based on each of the three models.  The $K$-band luminosity
gap was computed for each halo within a projected cluster-centric
radius of $640\kpc$.  The predicted luminosity gap statistic
distributions are over-plotted on the observed distribution in
Fig~\ref{fig:dm12histsim}.

The observed $\dm$ distribution is consistent, within the
uncertainties with a monotonically declining function of $\dm$
(\S\ref{sec:lumgap}).  The \citeauthor{bower06} model matches this
observational result well, and the predicted fraction of clusters with
the most extreme luminosity gaps is $p(\dm\ge2)=0.02^{+0.02}_{-0.01}$,
just $\sim1.3\sigma$ below the observed fraction of
$p(\dm\ge2)=0.07^{+0.05}_{-0.03}$ (\S\ref{sec:lumgap}).  In contrast,
the \citeauthor{croton06} model peaks at $\dm\sim1-1.5$ -- i.e.\ it
does not predict a monotonic decline of $p(\dm)$ -- however it
predicts $p(\dm\ge2)=8.6^{+2.4}_{-2.0}\%$ which is in excellent
agreement with the observations.  The \citeauthor{delucia07} model
predicts a yet more prominent peak at $\dm\sim1-1.5$, and a yet higher
fraction of clusters with extreme $\dm$,
$P(\dm\ge2)=16.7^{+4.5}_{-3.8}\%$, that disagrees with the
observations at $\sim2\sigma$.

Following the same approach as in \S\ref{sec:lumgap}, we also
decompose the predicted $\dm$ distributions into the predicted
absolute magnitudes of the first ($M_{K,1}$) and second ($M_{K,2}$)
ranked galaxies (Fig.~\ref{fig:m1dm12sim}).  The most striking feature
of this figure is that the slopes of $M_{K,1}$ versus $\dm$ and
$M_{K,2}$ versus $\dm$ are much steeper and shallower than the
observations respectively in the \citeauthor{bower06} model.  In
contrast, the \citeauthor{croton06} and \citeauthor{delucia07} models
succeed much better in reproducing the observed trends.
Interestingly, the discrepant trends in $M_{K,1}-\dm$ and
$M_{K,2}-\dm$ within the \citeauthor{bower06} model conspire to
produce a distribution of $\dm$ in Fig.~\ref{fig:dm12histobs} that is
in good agreement with observations.

The absolute magnitudes of BCGs span $\sim2\mags$ in the
\citeauthor{bower06} model, in contrast to the observed range of
$\sim1\mags$.  As BCGs grow, the largest increase in luminosity from
purely ingesting another galaxy is a brightening by $0.75\mags$, i.e.\
a merger between the brightest two galaxies in a cluster with $\dm=0$.
The very large spread in $M_K$ for BCGs in the \citeauthor{bower06}
model therefore indicates that the conversion of cold gas into stars
is too efficient in their model.  In the model, most of the mergers
that form BCGs are between gas poor galaxies.  The main source of gas
for formation of new stars is that which cools from the intracluster
medium.  The steep relationship between $M_{K,1}$ and $\dm$ therefore
implies that AGN feedback in BCGs is too weak in the
\citeauthor{bower06} model.  An important caveat on this
interpretation is that we showed in
\S\S\ref{sec:isophot}~\&~\ref{sec:summary} that clusters with large
luminosity gaps ($\dm\ge1$) have non-boxy isophotes and therefore
likely formed from mergers of gas rich galaxies, i.e.\ probably at
higher redshift than the BCGs in the model.

The shallow slope of the relationship between $M_{K,2}$ and $\dm$ in
the \citeauthor{bower06} model implies that the replenishment of the
supply of cluster galaxies that are ingested into their respective
BCGs is too efficient in this model.  Specifically, the difference in
slopes of $M_{K,2}-\dm$ between \citeauthor{bower06} and the other two
models could arise from differing treatments of the merging of
galaxies in the respective models following the time at which
individual galaxy halos lose their identity following ingestion into
the parent cluster halo.  We also comment, more generally, that the
galaxies in \citeauthor{bower06}'s model tend to be less luminous than
the observed galaxies by $\sim0.5\,\mags$, and those in
\citeauthor{croton06}'s model tend to be over-luminous by
$\sim0.3\,\mags$.  This suggests that the strength of feedback in the
general cluster population may be too strong in the former and too
weak in the latter model.

For completeness, we also compare our measurement of the fraction of
$10^{15}{\Msol}$ clusters that satisfy $\dm\ge2$ with predictions from
\citeauthor{milos06}'s (\citeyear{milos06}) analytic model.  Our
measurement of $0.07^{+0.05}_{-0.03}$ is well within $2\sigma$ of
\citeauthor{milos06}'s prediction of $0.03$.  The most obvious
difference between their model and our observations is that the
prediction is calculated within the cluster virial radii, in contrast
to our calculation within a projected cluster-centric radius of
$\sim0.4r_{200}$.  The larger volume within each cluster probed by
\citeauthor{milos06} will reduce the probability of finding clusters
with large luminosity gap statistics.  The same authors also estimate
the fraction of $10^{15}\Msol$ clusters with $\dm\ge2$ using data from
the Sloan Digital Sky Survey \citep{miller05}, obtaining a similar
fraction to their prediction.  The possible disagreement between this
estimate and our own is harder to understand because both use a
similar physical aperture for the calculation of $\dm$.  We note,
however, that the two observed cluster samples are selected in
different ways; our sample is X-ray selected whilst SDSS is optically
selected.

\section{Conclusions}\label{sec:conclusions}

We have combined wide-field near-infrared imaging from the WIRC camera
on the Hale 200in telescope, with \emph{HST}, \emph{Chandra}, and
\emph{Spitzer} observations of 59 massive galaxy clusters at
$z\simeq0.2$ to explore the connections between the formation
histories of BCGs and the galaxy clusters that they inhabit.  This
large statistical sample is intended to be representative of the
underlying population of massive X-ray luminous clusters.  Extensive
tests confirm that results based on this sample can be regarded as
statistically compatible with those from a complete volume-limited
sample.  Our main empirical results are as follows:

\smallskip

\noindent (i) We have made the first observational measurement of the
distribution of the luminosity gap statistic, $\dm$, of massive
$\sim10^{15}\Msol$ clusters.  The probability distribution of the
luminosity gap statistic is a monotonically declining function of
$\dm$, well described by the relation $p(\dm)=A+B\dm$ with
$A=0.41\pm0.03$ and $B=-0.13\pm0.02$.

\smallskip

\noindent (ii) Following \citet{dariush07} we used Monte Carlo
simulations to quantify the fraction of clusters with large luminosity
gaps expected from random sampling of a Schechter function.  The
observed distribution exceeds the statistical distribution derived
from the Monte Carlo simulation at $\dm\ge1$ at $\sim3\sigma$
significance, confirming that the most extreme luminosity gaps have a
physical origin, and are not statistical flukes.  

\smallskip

\noindent(iii) Four of our sample of 59 clusters have extreme
luminosity gaps of $\dm\ge2$ -- ZwCl\,1309.1$+$2216, A\,1835, A\,2261,
and RXC\,J2102.1$-$2431 -- which equates to a fraction of
$10^{15}\Msol$ clusters that have $\dm\ge2$ of
$p(\dm\ge2)=0.07^{+0.05}_{-0.03}$.

\smallskip

\noindent (iv) The morphology of 45/59 BCGs was measured by analyzing
the shape of the BCG isophotes in archival and new \emph{HST}
observations of the cluster cores.  The split between boxy, elliptical
and disky isophotes is: 22\% boxy, 32\% elliptical, 29\% disky, with
17\% unclassified.

\smallskip

\noindent (v) A strong correlation is found between $\dm$ and
$\fsub$, the fraction of mass in the cluster cores associated with
group- and galaxy scale dark matter halos, the latter coming from
published gravitational lens models of the cluster cores
\citep{smith05}.  The relationship between $\dm$ and $\fsub$ is
parameterized thus: $\log\fsub=\mu+\nu\dm$, with best fit parameters
$\mu=-0.29\pm0.15$ ad $\nu=-0.58\pm0.11$. 

\smallskip

\noindent (vi) Clusters with large luminosity gaps, $\dm\gs1-1.5$,
have cuspy gas density profiles, and thus relatively strong strong
cool cores ($\alpha\le-0.6$, where $\alpha$ is the logarithmic gas
density profiles at $0.04r_{500}$), elliptical or disky BCGs
($\mean{B_4}\ge0$, where $B_4$ is the fourth-order Fourier
co-efficient of the optical isophotes), concentrated dark matter
density profiles ($c_{500}\gs1$, where $c_{500}$ is based on a
\citealt{hernquist90} model fit to the \emph{Chandra} data), and small
substructure fraction ($\fsub\ls0.1$, where $\fsub$ is based on strong
lens modeling of the mass distribution).

\smallskip

\noindent(vii) In contrast, clusters with small luminosity gaps,
$\dm\ls1$, span the full range of observed cool core strengths
($-1.3\ls\alpha\ls0$), span the full range of boxy, elliptical, and
disky BCG morphologies ($-0.015\ls\mean{B_4}\ls-0.015$), span the full
range of concentrations ($c_{500}\sim0-2.5$), and have large
substructure fractions ($\fsub\gs0.1$).

\smallskip

Clusters with $\dm\gs1$ are therefore a more homogeneous population
than clusters with $\dm\ls1$.  The stronger cool cores, more
concentrated mass distribution, and non-boxy BCGs, all point towards
high-$\dm$ clusters forming at early times.  Such early formation is
required to allow sufficient time to pass for the BCG to ingest (aided
by dynamical friction) the bright cluster galaxy population in order
to develop the large luminosity gap, and for the establishment of the
cool core.  The formation of more concentrated dark matter halos at
earlier times than less-concentrated halos is a generic prediction of
cold dark matter theory \cite[e.g.][]{neto07}.  The interpretation of
disky BCGs is less straightforward, however such morphologies can
plausibly be interpreted as evidence for the last major mergers in a
BCG's formation history comprising gas-rich galaxies -- the presence
of gas thus leading to the establishment of a disk-like structure in
the BCG.  This gas-rich merger scenario for BCG formation is
consistent with the early formation of large-$\dm$ clusters.

How can the heterogeneous population of low-$\dm$ clusters, and more
specifically, the fact that some low-$\dm$ clusters have strong cool
cores, non-boxy BCGs, and high concentrations, be interpreted within
the context of the early formation of high-$\dm$ clusters?  The most
natural explanation is that large-$\dm$ clusters can evolve into
low-$\dm$ clusters when the supply of bright cluster galaxies is
replenished by episodes of hierarchical infall of smaller galaxy
systems, such as galaxy groups.  Such infall would depress $\dm$ and
increase $\fsub$ immediately that the group entered the measurement
aperture (in this case a clustercentric radius of $\sim0.4r_{200}$),
and would modify other cluster properties such the cool core strength,
BCG morphology, and concentration of the mass distribution on longer
timescales of several Gyr.  The observed heterogeneity of low-$\dm$
clusters can therefore be explained by these clusters comprising both
(i) clusters that have formed more recently, and thus have a low
concentration, haven't had time to develop a large luminosity gap and
cool core, and have a BCG formed from relatively gas-poor mergers, and
(ii) clusters that formerly had a large luminosity gap, and have
suffered hierarchical infall in the previous few Gyr.  We therefore
conclude that a large luminosity gap (and large substructure fraction)
is a phase through which a cluster can evolve if sufficient time
elapses between episodes of hierarchical merging of other galaxies and
groups of galaxies with the cluster.  The large scatter seen in the
theoretical age-$\dm$ and age-$\fsub$ relationships
\citep{dariush07,dariush10,smith08} lend further weight to the view
that both the age \emph{and} the recent merger history of a cluster
contribute to the observed values of $\dm$ and $\fsub$.

We also compare our observational results with predictions from
Millennium simulation-based semi-analytic models of galaxy evolution.
We find that none of the models can successfully reproduce the
observations in their entirety.  \cite{bower06} succeeds best at
reproducing the monotonically declining $p(\dm)$, however they predict
a relationship between BCG luminosity and $\dm$ that is far too steep.
In contrast, both \cite{croton06} and \cite{delucia07} predict that
$p(\dm)$ peaks at $\dm\sim1-1.5$, in disagreement with the
observations, with \citeauthor{delucia07} predicting the more
prominent peak.  \citeauthor{delucia07} also predict
$p(\dm\ge2)\sim0.17$, in contrast to the observed value of
$p(\dm=0.07^{+0.05}_{-0.03}$.  Nevertheless, both
\citeauthor{croton06} and \citeauthor{delucia07} match the observed
slope of the relationship between BCG luminosity and $\dm$ very well.
We discuss the possible causes of these disagreements, and suggest
that \citeauthor{bower06}'s model may be too efficient at converting
cold gas to stars in BCGs, and may also to be too efficient at
replenishing the supply of galaxies in clusters.

We also note that semi-analytic galaxy evolution models also fail to
reproduce observational results on high redshift BCGs
\citep{collins09,stott10}.  Our new results add to this picture of the
inability of models to reproduce observations of BCGs.  An important
strength of our results is that we do not rely on calculations of the
stellar mass of BCGs, and thus are insensitive to possible systematic
uncertainties in stellar mass estimates for observed BCGs arising from
alternative stellar population models.

Our future work on the hierarchical assembly of clusters at
$z\simeq0.2$ will take advantage of the wide-field multi-wavelength
dataset that we are assembling, including mid/far-IR observations with
\emph{Spitzer} and \emph{Herschel}, joint strong/weak-lens modeling of
the cluster mass distributions, our spectroscopic redshift survey of
cluster galaxies with MMT/Hectospec, and X-ray observations with
\emph{XMM-Newton} and \emph{Chandra}.

\section*{Acknowledgments}
We acknowledge helpful comments from the anonymous referee.  We thank
our LoCuSS collaborators, in particular Alastair Edge, Victoria
Hamilton-Morris, Jean-Paul Kneib, Yuying Zhang, and Nobuhiro Okabe,
for encouragement, assistance and many stimulating discussions.  HGK,
GPS, AJRS, TJP, and JPS acknowledge support from PPARC and latterly
from STFC.  GPS acknowledges support from the Royal Society.  GPS
thanks Andrew Benson, Richard Bower, Gabriella de~Lucia, and Malcolm
Bremer for helpful discussions and comments; Kevin Bundy, Brad Cenko,
Chris Conselice, Richard Ellis, Avishay Gal-Yam, Sean Moran, David
Sand and Keren Sharon for assistance with acquiring some of the
near-infrared data presented in this article; and Rick Burruss and
Jeff Hickey for their support at Palomar Observatory.

\bsp

\label{lastpage}

\begin{thebibliography}{99}
\bibitem[\protect\citeauthoryear{Allen}{2003}]{allen03} Allen S. W.,
    2003, Ap\&SS, 285, 247
\bibitem[\protect\citeauthoryear{Ascasibar \&
    Diego}{2008}]{ascasibar08} Ascasibar Y., Diego J.~M., 2008, MNRAS,
    383, 369
\bibitem[\protect\citeauthoryear{Bender}{1988}]{bender88} Bender,
    R. 1988, A\&A, 193, 7
\bibitem[\protect\citeauthoryear{Bender \etal}{1989}]{bender89} Bender
    R., Surma P., Doebereiner S., Moellenhoff C., Madejsky R., 1989,
    A\&A, 217, 35
\bibitem[\protect\citeauthoryear{Bertin \& Arnouts}{1996}]{bertin96}
    Bertin, E. and Arnouts, S., 1996, A\&A, 117, 393
\bibitem[{{Bildfell} {et~al.}(2008){Bildfell}, {Hoekstra}, {Babul},
  {Mahdavi}}]{Bildfell08} 
  {Bildfell}, C. and {Hoekstra}, H. and {Babul}, A. and {Mahdavi}, A.,
  2008, MNRAS, 389, 1637 
\bibitem[\protect\citeauthoryear{B\"{o}hringer
    \etal}{2004}]{bohringer04} B\"{o}hringer H.; Schuecker, P.;
  Guzzo, L.; et al., 2004, A\&A, 425, 367
\bibitem[\protect\citeauthoryear{Bower \etal}{2006}]{bower06} Bower
  R.\ G.\ \etal MNRAS, 370, 645
\bibitem[\protect\citeauthoryear{Broadhurst
    \etal}{2005}]{broadhurst05} Broadhurst T., Takada M., Umetsu K.,
  \etal, 2005, ApJ, 619, 143
\bibitem[\protect\citeauthoryear{Bullock \etal}{2001}]{bullock01}
  Bullock J.S., Kolatt T.S., et al., 2001, MNRAS, 321, 559
\bibitem[\protect\citeauthoryear{Buote \etal}{2007}]{buote07} {Buote}
  D.~A., {Gastaldello} F., {Humphrey} P.~J., {Zappacosta} L.,
  {Bullock} J.~S., {Brighenti}, F., {Mathews} W.~G., 2007, ApJ, 664,
  123 
\bibitem[\protect\citeauthoryear{Cole \etal}{2001}]{Cole01}
    {Cole}, S., {Norberg}, P., {Baugh}, C.~M., {Frenk}, C.~S.,
    {Bland-Hawthorn}, J., {Bridges}, T., \& {Cannon}, R. 2001, MNRAS,
    326, 255  
\bibitem[\protect\citeauthoryear{Collins \etal}{2009}]{collins09}
  {Collins}, C.~A. and {Stott}, J.~P. and {Hilton}, M. and {Kay},
  S.~T. and {Stanford}, S.~A. and {Davidson}, M. and {Hosmer}, M. and
  {Hoyle}, B. and {Liddle}, A. and {Lloyd-Davies}, E. and {Mann},
  R.~G. and {Mehrtens}, N. and {Miller}, C.~J. and {Nichol}, R.~C. and
  {Romer}, A.~K. and {Sahl{\'e}n}, M. and {Viana}, P.~T.~P. and
  {West}, M.~J., 2009, Nature, 458, 603
\bibitem[\protect\citeauthoryear{Crawford \etal}{1999}]{crawford99}
  {Crawford}, C.~S. and {Allen}, S.~W. and {Ebeling}, H. and {Edge},
  A.~C. and {Fabian}, A.~C., 1999, MNRAS, 306, 857
\bibitem[\protect\citeauthoryear{Croton \etal}{2006}]{croton06} Croton
  D.\ J.\ \etal, 2006, MNRAS, 365, 11
\bibitem[\protect\citeauthoryear{Cypriano \etal.}{2006}]{cyp06}
  Cypriano E. S., Mendes de Oliveira C., Sodre Jr. L., 2006, AJ, 132,
  514
\bibitem[\protect\citeauthoryear{Dariush \etal}{2007}]{dariush07}
  Dariush A. A., Khosroshahi H. G., Ponman T. J., Pearce F.,
  Raychaudhury S., Hartly W., 2007, MNRAS, 382, 433
\bibitem[\protect\citeauthoryear{Dariush}{2009}]{dariush09phd} Dariush
  A.~A., 2009, PhD Thesis, University of Birmingham, England
\bibitem[\protect\citeauthoryear{Dariush \etal}{2010}]{dariush10}
  Dariush A.~A., Raychaudhury S., Ponman T.~J., Khosroshahi H.~G.,
  Benson A.~J., Bower R.~G., Pearce F., 2010, MNRAS, in press,
  arXiv:1002.4414 
\bibitem[\protect\citeauthoryear{De~Propris \etal}{1999}]{depropris99}
  de Propris, R., 
    Stanford, S.\ A., Eisenhardt, P.\ R., Dickinson, M., Elston, R.,
    1999, AJ, 118, 719
\bibitem[{{Dolag} {et~al.}(2004){Dolag}, {Bartelmann}, {Perrotta},
  {Baccigalupi}, {Moscardini}, {Meneghetti}, \& {Tormen}}]{Dolag04} 
  {Dolag}, K., {Bartelmann}, M., {Perrotta}, F., {Baccigalupi}, C.,
  {Moscardini}, L., {Meneghetti}, M., \& {Tormen}, G. 2004, A\&A, 416,
  853 
\bibitem[\protect\citeauthoryear{Duffy \etal}{2008}]{duffy08} Duffy
    A.~R., Schaye J., Kay S.~T., Dalla Vecchia C., 2008, MNRAS, 390,
    64
\bibitem[\protect\citeauthoryear{Ebeling \etal}{1998}]{ebeling98}
    Ebeling H.; Edge, A. C.; Bohringer, H.; Allen, S. W.; Crawford,
    C. S.; Fabian, A. C.; Voges, W.; Huchra, J. P. 1998, MNRAS, 301,
    881
\bibitem[\protect\citeauthoryear{Ebeling \etal}{2000}]{ebeling00}
    Ebeling, H.; Edge, A. C.; Allen, S. W.; Crawford, C. S.; Fabian,
    A. C.; Huchra, J. P. 2000, MNRAS, 318, 333
\bibitem[{{Edge} {et~al.}(1999){Edge}, {Ivison}, {Smail}, {Blain}, \&
  {Kneib}}]{Edge99}{Edge}, A.~C., {Ivison}, R.~J., {Smail}, I.,
  {Blain}, A.~W., \& {Kneib}, J.-P., 1999, MNRAS, 306, 599
\bibitem[\protect\citeauthoryear{Edge \etal}{2003}]{edge03}
    Edge, A. C., Smith G.~P., Sand D.~J., Treu T., Ebeling H., Allen
    S.~W., van Dokkum P.~G., 2003, ApJ, 599, 69
\bibitem[{{Egami} {et~al.}(2006){Egami}, {Misselt}, {Rieke}, {Wise},
  {Neugebauer}, {Kneib}, {Le Floc'h}, {Smith}, {Blaylock}, {Dole}, {Frayer},
  {Huang}, {Krause}, {Papovich}, {P{\'e}rez-Gonz{\'a}lez}, \&
  {Rigby}}]{Egami06}{Egami}, E., {Misselt}, K.~A., {Rieke}, G.~H.,
  {Wise}, M.~W., {Neugebauer}, G., {Kneib}, J., {Le Floc'h}, E.,
  {Smith}, G.~P., {Blaylock}, M., {Dole}, H., {Frayer}, D.~T.,
  {Huang}, J., {Krause}, O., {Papovich}, C., {P{\'e}rez-Gonz{\'a}lez},
  P.~G., \& {Rigby}, J.~R. 2006, ApJ, 647, 922 
\bibitem[\protect\citeauthoryear{Evrard \etal}{2002}]{Evrard02}
  {Evrard}, A.~E. and {MacFarland}, T.~J. and {Couchman}, H.~M.~P. and
  {Colberg}, J.~M. and {Yoshida}, N. and {White}, S.~D.~M. and
  {Jenkins}, A. and {Frenk}, C.~S. and {Pearce}, F.~R. and {Peacock},
  J.~A. and {Thomas}, P.~A., 2002, ApJ, 573, 7
\bibitem[\protect\citeauthoryear{Faber \etal}{1997}]{faber97} Faber
  S. M., et al., 1997, AJ, 114, 1771
\bibitem[\protect\citeauthoryear{Gavazzi}{2003}]{gavazzi03} Gavazzi
  R., Fort B., et al., 2003, A\&A, 403, 11
\bibitem[\protect\citeauthoryear{Gehrels}{1986}]{gehrels86} Gehrels
  N., 1986, ApJ, 303, 336
\bibitem[\protect\citeauthoryear{Geller et al.}{1976}]{Geller76}
  {Geller}, M.~J. and {Peebles}, P.~J.~E., 1976, ApJ, 206, 939
\bibitem[\protect\citeauthoryear{Gratton}{1997}]{gratton97} {Gratton},
  R.~G. and {Fusi Pecci}, F. and {Carretta}, E. and {Clementini},
  G. and {Corsi}, C.~E. and {Lattanzi}, M., 1997, ApJ, 491, 749
\bibitem[\protect\citeauthoryear{Hamilton-Morris
    \etal}{2008}]{hamilton08} Hamilton-Morris V.~H., et al, 2009,
  in preparation
\bibitem[\protect\citeauthoryear{Haines \etal}{2009a}]{haines09a} Haines
  C.~P., Smith G.~P., et al., 2009, MNRAS, 396, 1297
\bibitem[\protect\citeauthoryear{Haines \etal}{2009b}]{haines09b} Haines
  C.~P., Smith G.~P., et al., 2009, ApJ, 704, 126
\bibitem[\protect\citeauthoryear{Haines \etal}{2010}]{haines10} Haines
  C.~P., Smith G.~P., et al., 2010, A\&A, in press, arXiv:1005.3811
\bibitem[\protect\citeauthoryear{Heckman}{1981}]{heckman81}
  {Heckman}, T.~M., 1981, ApJ, 250, 59
\bibitem[\protect\citeauthoryear{Hernquist}{1990}]{hernquist90}
  Hernquist L., 1990, ApJ, 356, 359
\bibitem[\protect\citeauthoryear{Jee \etal}{2007}]{jee07} Jee, M. J.;
  Ford, H. C.; Illingworth, G. D.; White, R. L.; Broadhurst, T. J.;
  Coe, D. A.; Meurer, G. R.; van der Wel, A.; Benítez, N.;
  Blakeslee, J. P., 2007, 2007, ApJ, 661, 728
\bibitem[\protect\citeauthoryear{Jing}{2000}]{jing00} Jing Y. P.,
  2000, ApJ, 535, 30
\bibitem[\protect\citeauthoryear{Jing \& Suto}{2002}]{jingsuto02}
  Jing, Y. P.; Suto, Y., 2002, ApJ, 574, 538
\bibitem[\protect\citeauthoryear{Jones \etal}{2003}]{jones03} Jones
  L. R., Ponman T. J., Horton A., Babul A., Ebeling H., Burke D. J.,
  2003, MNRAS, 343, 627
\bibitem[\protect\citeauthoryear{Jones \etal}{2000}]{jones00} Jones
  L. R., Ponman T. J., Forbes D.A., 2000, MNRAS, 312, 139
\bibitem[\protect\citeauthoryear{Khochfar \&
    Burkert}{2005}]{khochfar05} Khochfar S., Burkert A., 2005, MNRAS,
  359, 1379
\bibitem[\protect\citeauthoryear{Khosroshahi \etal}{2004}]{krpmf04}
  Khosroshahi H. G., Raychaudhury S., Ponman T. J., Miles
  T. A. Forbes D., 2004, MNRAS, 349, 524
\bibitem[\protect\citeauthoryear{Khosroshahi \etal}{2004}]{kjp04}
  Khosroshahi H.~G., Jones L.~R., Ponman T.~J., 2004, MNRAS, 349,
  1240
\bibitem[\protect\citeauthoryear{Khosroshahi \etal}{2006}]{kmpj06}
  Khosroshahi H.~G., Maughan B., Ponman T.~J., Jones L.~R., 2006,
  MNRAS, 369, 1211
\bibitem[\protect\citeauthoryear{Khosroshahi, Ponman \&
    Jones}{2006}]{kpj06} Khosroshahi H. G., Ponman T. J., Jones L. R.,
  2006, MNRAS Letters, 372, 68
\bibitem[\protect\citeauthoryear{Khosroshahi, Ponman \&
    Jones}{2006}]{kpj07} Khosroshahi H. G., Ponman T. J., and Jones
  L. R., 2006, MNRAS, 377, 595
\bibitem[\protect\citeauthoryear{King \& Ellis}{1985}]{king85} King
  C. R. and Ellis R. S., 1985, ApJ, 288, 456
\bibitem[\protect\citeauthoryear{Kneib \etal}{2003}]{kneib03} Kneib,
  J-P, \etal, 2003, ApJ, 598, 804
\bibitem[\protect\citeauthoryear{Komatsu \etal}{2009}]{komatsu09}
  Komatsu, E., et al.\ 2009, Astronomy, 2010, 158  
\bibitem[\protect\citeauthoryear{La~Barbera \etal}{2004}]{labarbera04}
  {La Barbera}, F. and {Merluzzi}, P. and {Busarello}, G. and
  {Massarotti}, M. and {Mercurio}, A., 2004, A\&A, 425, 797
\bibitem[\protect\citeauthoryear{La~Barbera \etal}{2010}]{labarbera10}
  {La Barbera}, F. and {de Carvalho}, R.~R. and {de la Rosa},
  I.~G. and {Gal}, R.~R. and {Swindle}, R. and {Lopes}, P.~A.~A.,
  2010, AJ, submitted, arXiv:1006.4065
\bibitem[\protect\citeauthoryear{Limousin \etal}{2007}]{limousin07}
  Limousin, M.; Kneib, J. P.; Bardeau, S.; Natarajan, P.; Czoske,
  O.; Smail, I.; Ebeling, H.; Smith, G. P., 2007, A\&A, 461, 881
\bibitem[\protect\citeauthoryear{Lin \etal}{2004}]{lin04a} Lin Y.-T.;
  Mohr, J.~J.; Stanford, S.~A., 2004, ApJ, 610, 745
\bibitem[\protect\citeauthoryear{Lin \& Mohr}{2004}]{lin04b} Lin
    Y.-T. \& Mohr J.~J., 2004, ApJ, 617, 879
\bibitem[\protect\citeauthoryear{de~Lucia \& Blaizot}{2007}]{delucia07} 
  de~Lucia G., Blaizot J., 2007, MNRAS, 375, 2
\bibitem[\protect\citeauthoryear{McCarthy \etal}{2008}]{mccarthy08}
    McCarthy I., Babul A., Bower R.~G., Balogh M., 2008, MNRAS, 386,
    1309
\bibitem[\protect\citeauthoryear{Mannucci \etal}{2001}]{mannucci01}
    Mannucci, F.; Basile, F.; Poggianti, et al., 2001, MNRAS, 
\bibitem[\protect\citeauthoryear{Mendes de Oliveira \etal}
  {2006}]{mendes06} Mendes de Oliveira C., Cypriano E. S., Sodre
  Jr. L., 2006, AJ, 131, 158 
\bibitem[\protect\citeauthoryear{Miller \etal}{2005}]{miller05} Miller
    C. J. \etal, 2005, AJ, 130, 968
\bibitem[\protect\citeauthoryear{Milosavljevi\'{c}
    \etal}{2006}]{milos06} Milosavljevi\'{c} M., Miller C. J.,
    Furlanetto S. R., Cooray A., 2006, ApJ, 637, L9
\bibitem[\protect\citeauthoryear{Marrone \etal}{2009}]{marrone09}
  Marrone D., Smith G.\ P., Richard J., et al., 2009, ApJ, 701, L114
\bibitem[\protect\citeauthoryear{Naab \& Burkert}{2003}]{naab03} Naab
    T., Burkert A., 2003, ApJ, 597, 893
\bibitem[\protect\citeauthoryear{Navarro \etal}{1997}]{nfw97} Navarro
    J. F., Frenk, C. S., White, S. D. M., 1997, ApJ, 490, 493
\bibitem[\protect\citeauthoryear{Neto \etal}{2007}]{neto07} Neto
    A.~F., Gao L., Bett P., et al., 2007, MNRAS, 381, 1450
\bibitem[\protect\citeauthoryear{Okabe \& Umetsu}{2008}]{okabe08}
    Okabe N., Umetsu K., 2008, PASJ, 60, 345
\bibitem[\protect\citeauthoryear{Okabe \etal}{2010}]{okabe09} Okabe
    N., Takada M., Umetsu K., Futamase T., Smith G.P., 2010, PASJ,
    in press, arXiv:0903.1103
\bibitem[\protect\citeauthoryear{Okabe \etal}{2010}]{okabe10} Okabe
  N., Zhang Y.-Y., Finoguenov A., Takada M., Umetsu K., Futamase T., 
  2010, ApJ, submitted
\bibitem[\protect\citeauthoryear{Oergerle \&
    Hoessel}{1989}]{Oergerle89} Oegerle, W.~R. and Hoessel, J.~G.,
  1989, AJ, 98, 1523
\bibitem[\protect\citeauthoryear{Oguri \etal}{2010}]{Oguri10}
  {Oguri}, M. and {Takada}, M. and {Okabe}, N. and {Smith}, G.~P.,
  2010, MNRAS, in press.
\bibitem[\protect\citeauthoryear{Ostriker \& Hausman}{1977}]
  {Ostriker77} Ostriker, J.~P. and Hausman, M.~A., 1977, ApJ, 217, 125
\bibitem[\protect\citeauthoryear{Pereira \etal}{2010}]{pereira10}
  Pereira M., Haines C.~P., et al., 2010, A\&A, in press,
  arXiv:1005.3813 
\bibitem[\protect\citeauthoryear{Ponman \etal}{1994}]{ponman94} Ponman
    T. J., Allan D. J., Jones L. R., Merrifield M., MacHardy I. M.,
    1994, Nature, 369, 462
\bibitem[{{Quillen} {et~al.}(2008){Quillen}, {Zufelt}, {Park}, {O'Dea}, {Baum},
  {Privon}, {Noel-Storr}, {Edge}, {Russell}, {Fabian}, {Donahue}, {Bregman},
  {McNamara}, \& {Sarazin}}]{Quillen08}
{Quillen}, A.~C., {Zufelt}, N., {Park}, J., {O'Dea}, C.~P., {Baum}, S.~A.,
  {Privon}, G., {Noel-Storr}, J., {Edge}, A., {Russell}, H., {Fabian}, A.,
  {Donahue}, M., {Bregman}, J.~N., {McNamara}, B.~R., \& {Sarazin}, C.~L. 2008,
  ApJS, 176, 39
\bibitem[\protect\citeauthoryear{Reid}{1997}]{reid97} {Reid}, I.~N.,
    1997, AJ, 114, 161
\bibitem[\protect\citeauthoryear{Reiprich \&
    B\"{o}hringer}{2002}]{reiprich02} Reiprich T. H. and B\"{o}hringer
    H., 2002, ApJ, 567, 716
\bibitem[\protect\citeauthoryear{Richard \etal}{2010a}]{richard09}
  Richard J., et al., 2010, MNRAS, 402, 44
\bibitem[\protect\citeauthoryear{Richard \etal}{2010b}]{richard10}
  Richard J., Smith G.\ P., et al., 2010, MNRAS, 405, 325
\bibitem[\protect\citeauthoryear{Sandage \& Hardy}{1973}]{sandage73}
  Sandage, A. and Hardy, E., 1973, ApJ, 183, 743
\bibitem[{{Sanderson} {et~al.}(2009){Sanderson}, {Edge}, \&
  {Smith}}]{sanderson09a}{Sanderson}, A.~J.~R., {Edge}, A.~C., \&
  {Smith}, G.~P. 2009, MNRAS, 398, 1698 
\bibitem[{{Sanderson} \& {Ponman}(2010)}]{Sanderson10}
  {Sanderson}, A.~J.~R. \& {Ponman}, T.~J. 2010, MNRAS, 402, 65
\bibitem[\protect\citeauthoryear{Smith \etal}{2005}]{smith05} Smith
  G. P. Kneib J., Smail I., Mazzotta P., Ebeling H., Czoske, O. ,
  2005, MNRAS, 359, 417
\bibitem[\protect\citeauthoryear{Smith \& Taylor}{2008}]{smith08}
  Smith G.~P., Taylor J.~E., 2008, ApJ, 682, L73
\bibitem[\protect\citeauthoryear{Smith \etal}{2009}]{smith09}
  Smith G.~P., et al., 2009, ApJ, 707, 163
\bibitem[\protect\citeauthoryear{Smith \etal}{2010}]{smith10} Smith
  G.~P., Haines C.~P., et al., 2010, A\&A, in press, arXiv:1005.3816
\bibitem[\protect\citeauthoryear{Stott \etal}{2008}]{stott08} Stott
  J.~P., Edge A.~C., Smith G.~P., Swinbank A.~M., Ebeling H., 2008,
  MNRAS, 384, 1502
\bibitem[\protect\citeauthoryear{Stott \etal}{2010}]{stott10} {Stott},
  J.~P. and {Collins}, C.~A. and {Sahlen}, M. and {Hilton}, M. and
  {Lloyd-Davies}, E. and {Capozzi}, D. and {Hosmer}, M. and {Liddle},
  A.~R. and {Mehrtens}, N. and {Miller}, C.~J. and {Romer}, A.~K. and
  {Stanford}, S.~A. and {Viana}, P.~T.~P. and {Davidson}, M. and
  {Hoyle}, B. and {Kay}, S.~T. and {Nichol}, R.~C., 2010, ApJ, in
  press, arXiv:1005.4681
\bibitem[\protect\citeauthoryear{Taylor \& Babul}{2004}]{taylor04}
  Taylor J.~E., Babul A., 2004, MNRAS, 348, 811
\bibitem[\protect\citeauthoryear{Vale \& Ostricker}{2007}]{vale07}
  Vale A., Ostriker J. P., 2007, astro-ph/0701096
\bibitem[\protect\citeauthoryear{Tremaine et
    al.}{1977}]{Tremaine77}{Tremaine}, S.~D. and {Richstone}, D.~O.,
  1977, ApJ, 212, 311 
\bibitem[{{Vikhlinin} {et~al.}(2007){Vikhlinin}, {Burenin}, {Forman}, {Jones},
  {Hornstrup}, {Murray}, \& {Quintana}}]{Vikhlinin07}
{Vikhlinin}, A., {Burenin}, R., {Forman}, W.~R., {Jones}, C., {Hornstrup}, A.,
  {Murray}, S.~S., \& {Quintana}, H. 2007, in Heating versus Cooling in
  Galaxies and Clusters of Galaxies, ed. {H.~B{\"o}hringer, G.~W.~Pratt,
  A.~Finoguenov, \& P.~Schuecker }, 48--+
\bibitem[\protect\citeauthoryear{Wechsler \etal}{2002}]{wechsler02}
    Wechsler R.~H., Bullock J.~S., et al., 2002, ApJ, 568, 52
\bibitem[\protect\citeauthoryear{Wilson \etal}{2003}]{wilson03}
    Wilson J.~C., et al., 2003, SPIE, 4841, 451
\bibitem[\protect\citeauthoryear{Zentner \etal}{2005}]{zentner05}
    {Zentner}, A.~R., {Berlind}, A.~A., {Bullock}, J.~S., {Kravtsov},
    A.~V., {Wechsler}, R.~H.
\bibitem[\protect\citeauthoryear{Zhang \etal}{2007}]{zhang07} Zhang
    Y.-Y., Finoguenov A., B\"{o}hringer H., Kneib J.-P., Smith G.~P.,
    Czoske O., Soucail G., 2007, A\&A, 467, 437
\bibitem[\protect\citeauthoryear{Zhang et al.}{2008}]{zhang08} Zhang
    Y.-Y., Finoguenov A., B\"{o}hringer H., Kneib J.-P., Smith G.~P.,
    Kneissl, R., Okabe, N., Dahle, H., 2008, A\&A, 482, 451
\end{thebibliography}
\end{document}